\RequirePackage{fix-cm}
\documentclass[smallextended]{svjour3}       
\smartqed  
\usepackage{graphicx}
\usepackage{natbib}
\usepackage{amssymb,amsmath}
\def\apjl {Ap.J. Letters}
\def\aplett {Ap. Letters}
\def\apj {Ap.J. }
\def\aj {A.J.}
\def\apjs {Ap.J. Suppl.}
\def\aaps {A\&A Suppl.}
\def\aap {A\&A }
\def\mnras {MNRAS }
\def\pasp {PASP}
\def\araa {Ann.Rev.A.A}
\def\nat {Nature}

\def\pasj {PASJ}

\def\ls{{_<\atop^{\sim}}}
\def\gs{{_>\atop^{\sim}}}
\begin{document}

\title{An Optical View of BL Lacertae Objects}
\subtitle{}

\titlerunning{BL Lacs}        

\author{Renato Falomo \and  Elena Pian \\ \and  Aldo Treves }


\institute{Renato Falomo \at
              INAF-Osservatorio Astronomico di Padova, Vicolo dell'Osservatorio 5, 35122 Padova, Italy \\
              Tel.: +39-049-8293464\\
              Fax: +39-049-8759840\\
              \email{renato.falomo@oapd.inaf.it}           
              \and
             Elena Pian\at
             INAF-Istituto di Astrofisica Spaziale e Fisica Cosmica, Via P. Gobetti 101, 40129 Bologna, Italy; 
             Scuola Normale Superiore, Piazza dei Cavalieri 7, 56126 Pisa, Italy;  INFN, Sezione di Pisa, Italy
           \and
           Aldo Treves\at
             Universit\`a dell'Insubria, Via Valleggio 11, 22100 Como, Italy; 
             INFN, Sezione di Milano Bicocca, Italy; ICRA
}

\date{Received: date / Accepted: date}
\maketitle

\begin{abstract}
BL Lac objects are active nuclei, hosted in massive elliptical galaxies, the emission of which  is  dominated by  a relativistic jet closely aligned with the line of sight.   
This implies the existence of a parent population of sources with a misaligned jet, that have been identified with low-power radiogalaxies.  The spectrum of BL Lacs,  dominated by non-thermal emission over the whole electromagnetic range, together with bright compact radio cores, high luminosities,  rapid and large amplitude flux variability at all frequencies and  strong  polarization make these sources an optimal  laboratory for high energy astrophysics.  A most distinctive characteristic of the class is  the weakness or absence of spectral lines, that  historically hindered the identification of their nature and ever thereafter proved to be a hurdle in the determination of their distance.   In this paper  we review the main observational facts that contribute to the present basic interpretation of this class of active galaxies.  We overview  the history of the BL Lac objects research field and  their population as it emerged from multi-wavelength surveys.  The properties of the flux variability and  polarization, compared with those at radio, X-ray and gamma-ray frequencies, are summarized together with  the present knowledge  of the host galaxies, their environments, and central black hole masses.  We focus this review on the optical observations, that played a crucial role in the early phase of BL Lacs studies, and, in spite of extensive radio, X-ray, and recently gamma-ray observations, could represent the future major contribution to the unveiling of the origin of these  sources.  In particular they could provide a firm conclusion on the long debated issue of the cosmic evolution of this class of active galactic nuclei and on the connection between formation of supermassive black holes and relativistic jets.

\end{abstract}


\vspace{1cm}

{\large \bf Preface}
\bigskip
\noindent

Discovered  and classified as an irregularly variable star by \cite{hoffmeister1929},  BL Lacertae was found to have a radio counterpart \citep{macland1968}.    This, together with the high optical polarization, drew attention on this source and on others that exhibited similar properties, primarily a highly variable and featureless, power-law-shaped optical spectral continuum.    This class of objects was soon recognized to be extragalactic in nature and to share some of the characteristics of quasars  \citep{strittmatter1972}.   Supporting evidence of this came from the detection of  a surrounding nebulosity centered on the optical point-like source  \citep[first noted by ][]{schmitt1968}, that hinted at the presence of a host galaxy \citep{ulrich78,ulrich89}.

The  peculiarities that make  BL Lac objects (BLLs) so enigmatic can be composed in a coherent picture that foresees a relativistic jet viewed at a small angle as the main responsible of all multi-wavelength BLL characteristics \citep{blarees1978}.  This explanation was later reinforced by the elaboration of a unifying scenario that envisages radiogalaxies, for which kiloparsec-scale jets are directly observed, and BLLs as members of the same population, with the viewing angle determining the different observed properties. 

Thanks to their widely extended spectral energy distribution, that covers sixteen decades in frequency, from the radio to TeV, BLLs can be detected with observations at any frequency.  In fact, surveys at radio, optical, X-ray and gamma-ray frequencies in the last 40 years  have contributed to select BLL candidates.  After identification via optical spectroscopy, samples of bona fide BLLs have been constructed at different frequencies and have made  the statistical study of the population possible.   A crucial parameter for assessing their cosmological role and evolution is their redshift, a quantity that is intrinsically difficult to determine in BLLs and for which accurate optical spectroscopy is critical.  By 1976 there were 30 known BLLs  \citep{stein76};  almost forty years on, there are nearly 1400 catalogued BLLs  \citep{vv2010}. 

Notwithstanding the importance of the multi-wavelength investigation, the observation and analysis of BLLs in the optical band remain a mainstay of research in this field, as witnessed by the advances in the last decade.  In this review  we have therefore focused on the optical domain. 
The first specific review  on BLLs was that of \cite{stein76}.   BLLs were the subject of three dedicated conferences:  at the University of Pittsburgh in 1978 (Pittsburgh Conference on BL Lac Objects, Proceedings edited by A.M. Wolfe, NSFP Pittsburgh, 1978),  in Como, Italy, in 1988 (BL Lac Objects, Eds. L. Maraschi, T. Maccacaro, M.-H. Ulrich, Lecture Notes in Physics, Springer-Verlag, 1989), and  in Turku, Finland, in 1998 (BL Lac phenomenon,  Eds. L. O. Takalo and A. Sillanp\"a\"a,  Astronomical Society of the Pacific, 1999).


\section{Introduction}
\label{sec:intro}

Galaxies are traditionally divided in active and inactive, according to whether their nuclei exhibit the telltale 
signs (broad emission lines, flux variability, radio compactness, strong emission at high energies) of a central 
``engine" or not.  However, the distinction is not sharp and there is a continuum of properties between these two 
extremes, with a wide range of nuclear activity levels, that accordingly define different types of active galactic 
nuclei (AGNs).  The most powerful among these are  luminous emitters over the whole electromagnetic 
spectrum up to the TeV energies, highly variable and highly polarized (bolometric luminosities can reach $10^{48}$ 
erg~s$^{-1}$ and are often dominated, especially during outbursts, by the gamma-ray emission).  These are called ``blazars"\footnote{This term  was coined by  E. Spiegel during the Conference on BL Lac Objects in Pittsburgh, in April 1978, from  a contraction of ``BL Lac" and ``quasar".}.

There is unanimous consensus on the fact that blazars owe their extreme physical behavior to the presence of a jet 
that is closely aligned with the observer's direction (estimated viewing angles $\theta \ls 10$ deg) where 
the plasma moves with a Lorentz\footnote{$\Gamma$ is defined as $(1 - \beta^2)^{-0.5}$, where $\beta = v/c$, with $v$ representing the plasma velocity.} factor $\Gamma$ of the order of $\sim 10$, and occasionally as high as $\sim$50 \citep{begfabrees2008}.  
The highly relativistic kinematic regime and  the small viewing angle produce a Doppler aberration (quantified by the factor $\delta = [\Gamma (1 - \beta  cos\theta)]^{-1}$), that foreshortens the observed time-scales, blue-shifts the observed spectrum and magnifies\footnote{In general the luminosity enhancement is a factor $\delta^p$, where $ 3 \ls p \ls 4$, depending on the geometry of the emitting region and on spectral index.} the luminosities at all wavelengths  \citep[e.g.][]{gg1993,urry95,gg2013}. Direct evidence for this interpretation comes from the detection of superluminal motion of plasma blobs along the jet, the  apparent velocity of which, as measured with radio interferometry, can be up to $20c$ \citep{marscher2008,jorstad2010}. 

When corrected for  Doppler aberration, these velocities are obviously 
lower than the speed of light, and compatible with a Lorentz factor of the order of 10.  Indirect arguments for relativistic beaming include: 
i) the super-Eddington accretion regimes for putative central black hole masses of $\sim 10^8$ M$_\odot$; 
ii) the transparency of the emitting regions to electron-positron pair-production 
at gamma-ray energies higher than $\sim$1 MeV  \citep{mcbreen1979,guilbert1983,maraschi1992}; 
iii) the huge radio brightness temperatures.  If no beaming is invoked then one  would expect  self-absorbed radio spectra 
over wide radio wavelength ranges and the Compton catastrophe in the X-rays. None of these effects are observed  \citep{hbs1966,woltjer1966,kellerptoth1969}.

The interpretation of the blazar class as relativistically beamed objects implies the existence of a parent population of misaligned sources.
Based on optical and radio properties  and population density, the most favored candidates are radiogalaxies\footnote{\cite{fanril1974}  
distinguished  radiogalaxies into two types: FR  I and FR II according to whether their luminosities at 178 MHz were lower or higher than $2 \times 10^{32}$ erg s$^{-1}$ Hz$^{-1}$sr$^{-1}$.  Their radio morphologies are also different, being core-dominated or lobe-dominated, respectively.
The  comparison of radio power and morphology and host galaxy luminosity of BLLs and radiogalaxies suggests that the dominant parent population consists of FR I radiogalaxies.}   that would be detected and classified as blazars when viewed at small angles  \citep{urry95}.

The term {\it blazar} encompasses bright radio-loud AGNs with compact radio cores, high amplitude multi-wavelength variability and large radio and optical polarization.  Historically,  those with weak or absent optical spectral  lines were named BL Lac objects (see also Section \ref{sec:specscopy}),  and the others received definitions that reflected their main properties: Optically Violently Variables, or Highly Polarized Quasars \cite[e.g., ][]{moore81}.   More recently, the distinction within the blazar class was drawn between objects with luminous broad emission lines, often accompanied also by prominent ultraviolet-optical continuum emission ({\it blue bump}) of thermal origin as normally seen in quasars, that are called Flat Spectrum Radio Quasars (FSRQ) and BLLs, where these broad emission lines are weak or absent \citep{urry95,padovani2012}.  This distinction is reflected also  in the radio polarization properties: at cm wavelengths BLLs tend to have polarization  vectors nearly parallel to the jet, while in FSRQs the vectors cover a wider range of directions, favoring  directions perpendicular to the jet \citep{gabuzda1989,marscher2002}.  
Owing to these different properties, and in  particular to the lack of  a substantial thermal component (accretion disk and/or dusty torus),  BLLs offer a  more 
direct view into the primary energy-production  mechanism, with respect to   FSRQs.

This review is organized as follows: in  Section \ref{sec:mwlref} we present the basic multi-wavelength properties and the reference model for BLLs; 
Section \ref{sec:demograph} reviews the methods of detection and selection of BLLs through surveys and the luminosity functions; 
Section \ref{sec:opticalvar} reports on the optical, near-infrared and ultraviolet variability of BLLs;  Section  \ref{sec:polarim} summarizes the optical polarization issues; the optical spectral properties, both from broad-band photometry and spectroscopy,  are described in Section \ref{sec:specscopy}; the observations of host galaxies are reviewed in Section \ref{sec:hostgals}; a few BLL sources exhibited optical counterparts to their radio jets,  as detailed in Section \ref{sec:mwljets}; the close environments and clustering properties of BLLs are reported in  Section \ref{sec:environs}; Section \ref{sec:bhmass} focuses on the  
central supermassive black hole, on its influence on the observed properties and on  its mass measurement.
Section \ref{sec:conclrem} contains a summary and future perspectives.


\section{Multi-wavelength properties  and the reference model}
\label{sec:mwlref}

BLLs are characterized by  strong non-thermal emission  at all frequencies.  In the radio band, they exhibit bright  \citep[up to $10^{31}$ erg~s$^{-1}$~Hz$^{-1}$, ][]{padovani2007}, compact cores with  power-law spectral continua with very flat shapes ($F_\nu \propto \nu^{-\alpha}$ with  $\alpha < 0.5$), mostly produced by  synchrotron radiation, often partially self-absorbed at the GHz frequencies\footnote{The overlap of several spatial emission components may also make the radio spectrum deviate from a Rayleigh-Jeans shape, so that it appears thin, although the individual components may be self-absorbed.}.

At infrared-optical wavelengths the spectrum of BLLs also follows a power-law, when  corrected for Galactic  extinction and host galaxy contribution.  The optical/near-infrared spectral indices are in the range $\alpha_\nu \sim 0.5-1.5$, i.e. they are steeper than in the radio \citep{falomo1993a,falomo1993d,pian1994}.  If the radio-to-optical spectrum is produced by the same optically thin synchrotron radiation component, the spectral steepening between these bands is naturally expected by increasing  synchrotron losses at higher frequencies 
 and adiabatic expansion of the jet plasma. The 
precise spectral slopes and their variability are also regulated by 
acceleration mechanisms and time-scales \citep{kirk1998,asano2014}.

In the X-rays BLLs are powerful sources ($L_{1 keV} \sim 10^{28}$ erg s$^{-1}$~Hz$^{-1}$), and in fact they were detected   by many satellites in the last 40 years  \citep{perlman1996,smu1996,wolter1998,pian1998,giommi1999,maselli2010}.  The X-ray spectra are generally well described by single power-laws or  curved slopes, depending on the  emission mechanism.  
If this is synchrotron radiation,  X-rays are generated by the highest energy electrons, so that the spectrum  has a relatively steep spectral index ($\alpha_\nu \sim 1.3$).  If the cooling  frequency (or break frequency)   of the synchrotron spectrum is located in the X-ray range, the spectrum has a convex shape in this  band. 
On the other hand, in a leptonic jet scenario (which we favour with respect to the hadronic one, see below), X-rays may be due to Compton up-scattering
of lower energy photons by the same electrons that produce the synchrotron spectrum (synchrotron self-Compton), so that the X-ray spectral shape is 
flat ($\alpha_\nu < 1$).   In BLLs  where the synchrotron component peaks below the X-ray range, the X-ray spectrum is occasionally concave 
and joins smoothly with the high energy tail of the observed synchrotron spectrum.

The  observations conducted in the MeV-GeV range from satellites (first the {\it 
Compton} Gamma-Ray Observatory in 1991-2000 and recently the orbiting {\it Fermi} Gamma-ray 
Satellite and {\it AGILE}) detected large numbers ($\sim$1000 to date) of extra-galactic 
sources that  turned out to be blazars, with very few exceptions 
\citep{hartman1999,ackermann2011,vercellone2011}.   A large subset of them are BLLs.  
The integrated luminosities in the range 0.1-10 GeV can reach $10^{49}$ erg s$^{-1}$.

Very high energy photons ($>$
100 GeV) from blazars were detected from the ground by atmospheric Cherenkov radiation 
telescopes starting about 20 years ago \citep{punch1992}.    About 50  TeV blazars are now 
known \citep[see e.g.][]{perdea2008,acciari2011,aliu2011, aliu2013,aleksic2011,aleksic2012,hesscoll2013,senturk2013},    many of   which 
have redshift higher than $\sim$ 0.2.  
A substantial fraction of these\footnote{see list at http://tevcat.uchicago.edu} are BLLs, especially  of high-frequency-peaked type, in which  TeV detection is favoured by the  broad-band spectral shape  (see below).
At $z \sim 1$ the extragalactic infrared background suppresses the TeV flux via pair production, so that  detection beyond this redshift is virtually impossible.
The highest observed luminosities at energies larger than 100 GeV are $\sim 10^{46}$ erg s$^{-1}$ \citep{magic2008}.

The  gamma-ray emission dominates the  spectral energy distribution of blazars in general, and BLL in particular,  especially during 
flaring states, and is highly variable  \citep[down to sub-hour time-scales, ][]{aharonian2009}.   In the leptonic scenario, this is
ascribed to inverse Compton scattering of lower energy photons off the most relativistic 
electrons.  Therefore the gamma-ray spectral shape is determined by the high energy tail of 
the electron distribution and by the shape of the spectrum of the scattered photons, which 
can be either optical-to-soft-X-ray synchrotron photons (synchrotron self-Compton) or photons 
external to the jet, like those from the accretion disk, broad-line region, dust torus 
(external Compton).

\begin{figure}
\centering
\includegraphics[width=0.9\columnwidth]{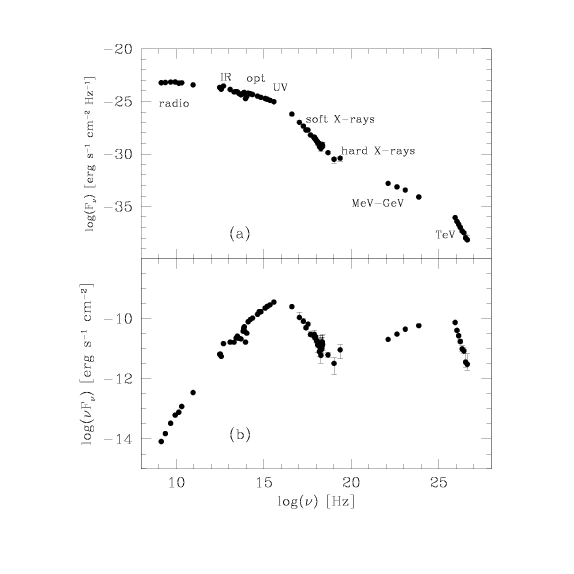}
\caption{Broad-band spectrum  of  PKS~2155--304 ($z = 0.116$).  Panel (a):
spectral flux distribution from radio to TeV frequencies; panel (b):   $\nu F_\nu$ representation of the spectrum,  that  emphasizes the frequencies at which most of the power is emitted.  Adapted from  \cite{foschini2007,hesscoll2012}.}  
\label{sed21552006}     
\end{figure}

In Figure \ref{sed21552006} we report an example of an overall spectrum of BLL.  
The spectral energy distribution has two humps, one peaking at the characteristic synchrotron cooling frequency (typically between the far-infrared and  soft  X-rays), 
and a second one attributed to inverse Compton cooling  in the MeV-GeV range.
These  frequencies vary from object to object  in a correlated way:   to a  larger frequency of the first hump corresponds a larger frequency of the second hump.
Accordingly,  BLLs  -- depending on their spectral energy distribution -- were traditionally divided in low-frequency-peaked objects (LBLs) and high-frequency-peaked objects  \cite[HBLs, ][]{padgio1995}.   Objects with intermediate properties are called intermediate BLLs (IBLs).  This distinction reflects broadly the previous concept of blazar classification according to radio or X-ray selection.  The characteristic spectral maxima in fact  favor naturally the detection of  a given blazar in  a specific band.  More extended multi-wavelength observations on large datasets led to a more complex picture (see Section \ref{sec:demograph}).

\begin{figure}
\centering
\includegraphics[width=0.9\columnwidth,height=0.5\textwidth]{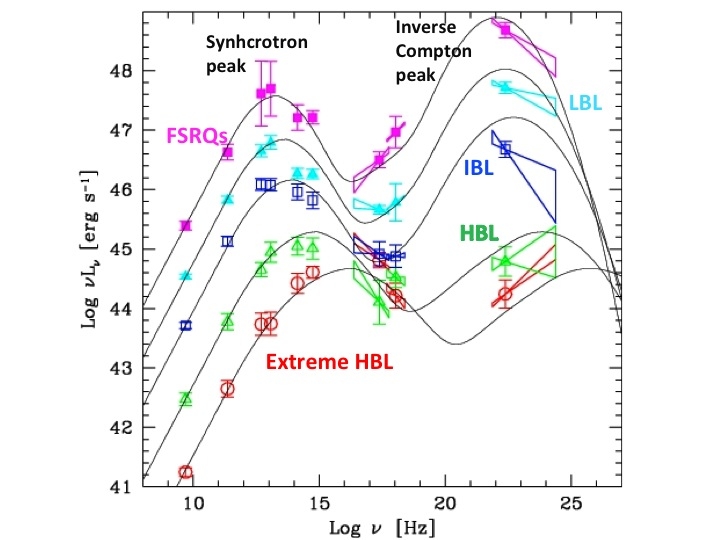}
\caption{Overall spectral energy distributions of blazars.  Note the differences of the relative intensities and frequencies of the two emission peaks  for various types of objects (see text).  This behavior is referred to as 
blazar sequence  \citep[from][]{fossati1998}.}
\label{figblazarseqprops}     
\end{figure}

In Figure \ref{figblazarseqprops} we show  spectral energy distributions of representative blazar sources.  While the double-humped shape is present in all of them,  the frequencies of the spectral peaks and their relative intensities differ significantly.  
\cite{fossati1998} found that the spectral energy distribution shapes  define a  continuum of properties,  whereby
(i) the first peak occurs in different
frequency ranges for different samples/luminosity classes, with most luminous sources
peaking at lower frequencies; 
(ii) the peak frequency of the gamma-ray component correlates
with the peak frequency of the lower energy one; 
(iii) the luminosity ratio between the high and
low frequency components increases with bolometric luminosity (Fig. \ref{figblazarseqprops}).  
In this picture,  called {\it blazar sequence},  BLLs are the sources with lower luminosities and higher characteristic synchrotron and inverse Compton frequencies.   The interpretation of this sequence is complex and somewhat controversial.   The parameters that govern it are the cooling efficiency of the relativistic particles, the accretion efficiency and ultimately the mass of the central black hole, BLLs being those with lower efficiencies and lower masses \citep{costamante2001,gg2008,gg2010}.   A number of objects deviate at first glance from this scheme, having comparatively high total luminosities and high synchrotron peak frequencies \citep{padovani2012,pottercotter2013}.

The central engine that powers the relativistic  jet is  assumed to be a rotating accreting black hole of about $10^8 M_\odot$. The mechanism of energy extraction and particle acceleration along the jet is however not completely understood  \citep[see e.g.][]{blazna1977,bbrrvmp1984,martav2003}.
The dynamics of the jet may be rather complex and it has in some cases been interpreted in terms of
helical structures tightly related to the black hole spin or to the orbital motion of a binary black hole system
\citep{camenkrock1992,ostorero2004}.   On sub-parsec scales, jets are pervaded by magnetic fields of the order of 0.1-10 Gauss that may have tangled geometries  and cause  particles --  leptons or hadrons -- to radiate through the synchrotron mechanism.

As seen above,  in the leptonic scenario  the main emission mechanisms are  synchrotron radiation of relativistic electrons at the  radio to X-ray wavelengths, which is responsible for the first hump of the spectral energy distribution,  and  Compton up-scattering of synchrotron photons (self-Compton)  or photon fields external to the jet at the higher energies, responsible for the second hump  \citep{gg1985,gg1998,blazejowski2000,fdb2008,pottercotter2013}.    For self-Compton scattering (that is the most common case in BLLs),  Klein-Nishina suppression \citep{moderski2005} may reduce the luminosity of the second hump and decrease its peak frequency.

In hadronic models, both primary electrons and protons are accelerated to ultrarelativistic energies, with
protons exceeding the threshold for photo-pion production on the soft photon field in the
emission region. While the low-frequency emission is still due to synchrotron emission
from primary electrons, the high-energy emission is dominated by proton synchrotron emission, neutral pion decay photons, synchrotron and Compton emission from secondary decay products
of charged pions, and the output from pair cascades initiated by these high-energy emissions
intrinsically absorbed by photon-photon pair production \citep{manbie1992,mannheim1998,aharonian2000,muecpro2001,muecke2003,boettcher2013}. 


\section{Demographics of BL Lacs}
\label{sec:demograph}

In order to understand the cosmological evolution of BLLs  in comparison with other AGN and derive the properties of the unbeamed parent population it is mandatory to construct statistically significant samples of sources with measured redshift. 
Statistical samples are ultimately extracted from unbiased  surveys carried out at various wavelengths.  Depending on the adopted   band,  objects identified as BLL candidates, and then confirmed through the essential  criterion of optical spectroscopy,  are  reported usually as radio, X-ray, optically, and also gamma-ray-selected objects (see Section \ref{sec:mwlref}).
The characteristic double-peaked spectral energy distribution of BLLs (see Fig. \ref{figblazarseqprops}) emphasizes the effects of single band flux-limited surveys.  As an example, radio surveys preferentially select the sources with a low-frequency-peaked synchrotron component.   On the other hand, in GeV surveys the objects with high-frequency inverse Compton component are favoured.

\subsection{Surveys}
\label{sec:surveys}

\begin{figure}[h]

\includegraphics[width=0.5\columnwidth]{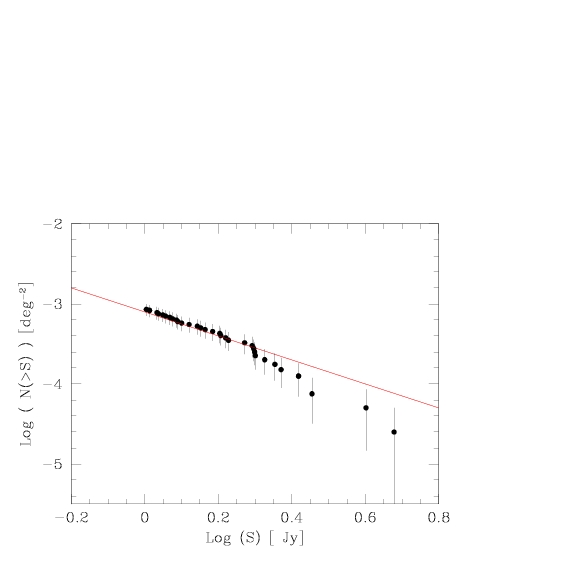} 
\includegraphics[width=0.5\columnwidth]{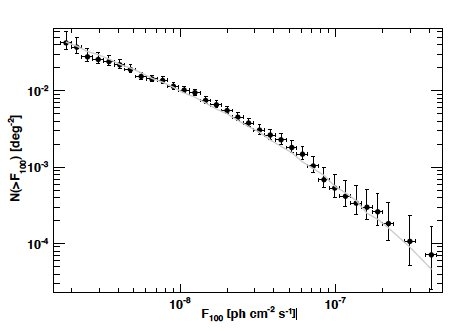}

\caption{Left:  Integral number counts for the 1 Jy BLLs  compared with the 
Euclidean  relationship (red solid line). Adapted from  \cite{stickel91}. Right: Integral number density of BLLs from Fermi 
\citep[from ][]{ajello2014}.}
\label{fig:ncounts}      
\end{figure}

The first complete flux-limited radio-selected sample of BLLs  was constructed from the 1 Jy catalogue of radio sources of  \cite{kuehr81} containing 518 sources with S$_\nu >$ 1 Jy at 5 GHz. The selection of BLLs  from this catalogue was
performed according to various properties (flat or inverted radio spectrum, optical magnitude and presence of weak
spectral lines). This selection produced a list of 34 radio-selected BLLs \citep{stickel91}.   The number counts derived from  this
sample (see Fig. \ref{fig:ncounts}, left)  showed a roughly Euclidean  behavior (no evolution), a result that was
then later confirmed by more extended and complete studies   \cite[e.g.][and  references therein]{caccianiga2008}. 

Deep and systematic surveys undertaken in the 80's with the then orbiting X-ray satellites discovered BLLs and thus contributed to the construction of complete samples of  BLLs selected on the basis of their X-ray emission.
From the Extended Medium Sensitivity Survey of the {\it Einstein} X-ray satellite
\citep[0.3--3.5 keV,   sensitivity $\sim 5 \times 10^{-14}$ erg~s$^{-1}$ cm$^{-2}$, ][]{gioia1990} \cite{stocke90}  extracted a  sample of 22 X-ray-selected BLLs.
The  {\it Einstein} Slew survey \citep{elvis1992} covered a much larger fraction of the sky  with shallower sensitivity ($\sim 5 \times 10^{-12}$ erg~s$^{-1}$ cm$^{-2}$).   A result of this survey is  a sample of 48 BLLs \citep{perlman1996}. 
Further X-ray surveys with {\it ROSAT} and {\it Swift} allowed the discovery of about 100 new BLLs \citep{voges1999,laurent1998,laurent1999,cusumano2010}.

In the optical band the selection of BLLs is more tricky.
The first BLLs discovered in the optical  were derived from the search of ultraviolet excess stellar objects  in the Palomar plates \citep{green1986}.  This resulted in the discovery of 7 objects \citep{fleming1993}.
Most fruitful recent optical  searches of BLLs resulted from the analysis of pure optical spectra of SDSS
sources.   \cite{plotkin2010}, starting from the 7th SDSS spectroscopic data base  \citep{abazajian2009} present a sample of $\sim$700 BLL  optically selected candidates.  A large fraction of the objects in the survey are of unknown redshift.   Search for radio counterparts of these candidates showed that  $\sim$90\% of these are radio-loud.   

Owing to the blazar spectral energy distribution, that extends to  very high energies, surveys in the GeV  band turned out to be extremely efficient in finding BLLs, thanks to their regular and uniform sky coverage.   In the 2nd Fermi LAT AGN Catalog \citep{ackermann2011},  which contains  $\sim$1000 sources, 80\% are blazars, of which about half are BLLs.  The majority of the remaining 20\% are blazar candidates.  In the selection of candidates,  near-infrared observations from the ground have been used in the past \citep[see e.g.][]{bersanelli1992,chen2005}, but  particularly effective were proven to be the spectra obtained from satellites   \citep[e.g. {\it Spitzer, }][]{chen2011}, and especially  the infrared colors from  \textit{WISE} observations, because gamma-ray blazars detected by {\it Fermi}  occupy a distinctive strip in the 3.4-4.6, 4.6-12 $\mu$m color-color diagram 
\citep{massaroe2012,massarof2013,dabrusco2013,paggi2013}.
These candidates must be eventually confirmed  via optical spectroscopy.

At variance with most AGNs, since BLLs are strong emitters at any wavelength, they can be detected in surveys performed in various spectral bands.  Moreover, owing to the high-amplitude flux variability they may appear in a survey depending on their activity state.  Modern BLL catalogs 
are based on multi-wavelength data and reflect  more accurately the full range of their spectral and timing properties \citep{schwku1983,perlman1996,maccacaro1998,giommi1999,beckmann2003,giommi2005,padovani2007,massaroe2009}.

\subsection{Luminosity Functions and parent population}
\label{sec:2.2}

The knowledge of the luminosity function of BLLs is instrumental to quantifying the beaming and thus to identifying the  parent population.  Its  construction  and  evolution (redshift dependence) are more complicated than for other classes of AGN, because of the difficulty of distance determination, and of the consequences of the radiation beaming  \citep{urry1984,maccacaro1984,urry1991,caccianiga2002,marcha2013}.
The luminosity functions of BLLs at radio and X-ray frequencies  had been evaluated, often with the purpose to compare them with those of radiogalaxies, in search of 
similarities that could prove the direct link between the two classes, under the assumption that they only differ in the viewing angle.  A common result of these efforts is that BLLs and FSRQs can be plausibly connected 
to FR  I and FR II radiogalaxies, respectively \citep{urry95,rector2000,chiaberge2001,capetti2002}, a conclusion that was independently reached also based on their similar host 
galaxies and environments (see also Sections \ref{sec:hostgals} and \ref{sec:environs}), emission line intensities, and radio morphologies 
\citep[for a review see ][]{marrov1994}. 
The detailed comparison of  radio and optical properties of BLLs and of their alleged parent  population (radiogalaxies) might require more complex jet geometries  \cite[e.g. structured jets,][]{ccc2000,trussoni2003}.

The partial and sometime contradictory results obtained from the analysis of luminosity functions are plagued by the limited size of  the original radio and X-ray samples \citep[see e.g.][]{morris1991,giommi1999,rector2000,caccianiga2002,padovani2007,marcha2013}.  The current surveys in gamma-rays yield much more numerous samples.
In their compilation of BLLs based on the first year of {\it Fermi} operations \citep{abdo2010a}  \cite{ajello2014} attempt to overcome the previous
limitations by including BLLs with spectroscopic redshifts and BLLs with matching redshift estimates from 
broad-band photometry and intervening absorption systems spectroscopy (Fig. \ref{fig:ncounts}, right).  
Using this complete sample of 211 
BLLs,   they construct  accurate gamma-ray luminosity functions of these sources.  Their tentative outcome is that positive evolution is  suggested for the majority of  sources at luminosities larger than $10^{45}$ erg~s$^{-1}$, while those at low luminosities may behave differently.

Their comparison with FSRQs \citep{ajello2012} suggests that the local ($z  \simeq 0$) luminosity function of BLLs overlaps and connects smoothly to 
that of FSRQs, highlighting the similarity between the two classes, with  BLLs having on average lower luminosity (and thus very likely lower Lorentz factors) 
than FSRQs.  \cite{ajello2014} also propose that BLLs  correspond to the final (gas-starved, inefficiently accreting) and long-lasting phase of 
an earlier, short-lived, merger-driven, gas-rich epoch,  represented by the FSRQs. A confirmation of this transition scenario can only rely on a 
more robust beaming correction and knowledge of the black hole masses and host 
galaxy environments, which are at present not well constrained.


\section{Flux variability}
\label{sec:opticalvar}

\begin{figure}
\centering
\includegraphics[width=0.9\columnwidth]{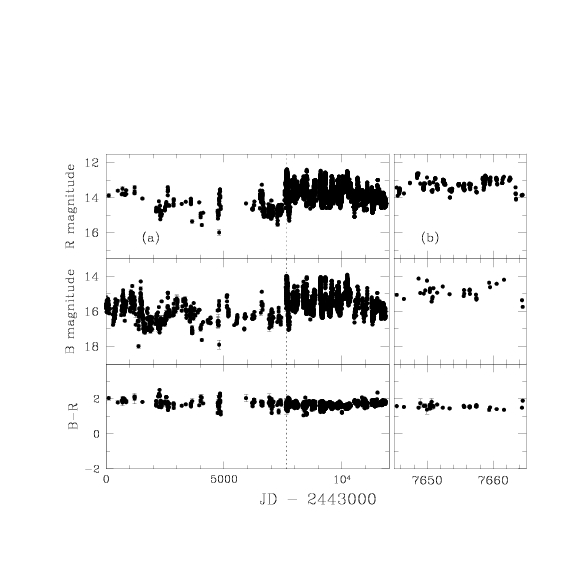}
\caption{(a)  Light curves of BL Lacertae in R filter (top), B filter (middle) and B -- R color (bottom), during $\sim$30 years since 1976.    Flux variations  of up to $\sim$3 magnitudes are observed, while  the color  varies within $\sim$1 magnitude.  (b) Expanded view of variability during the 3-weeks period   indicated as a vertical dashed line in panel (a).    Variations of the order of $\sim$1 magnitude are seen in a time-scale of few days, while the color remains  almost constant. The light curves were produced using data from the WEBT archive (courtesy of C.M. Raiteri and M. Villata).
}
\label{phvar_bllac}     
\end{figure}

One of the distinctive properties of BLLs is high amplitude flux variability (up to 5 magnitudes) over a range of time-scales from months down to minutes \citep{mcg1989,ww1995}.  Variability is present over the whole near-infrared-to-ultraviolet range and  does not exhibit periodic behavior; occasional outbursts of few magnitudes occurring in tens of days are observed.     It is generally achromatic, in particular for sources with their synchrotron peak located at frequencies higher than the optical (Fig. \ref{phvar_bllac}).

While in the majority of AGNs the variability time-scale can be almost directly linked to the size of the central supermassive black hole and  to the accretion rate \citep{mchardy2006}, in blazars this correlation does not hold because the luminosity is
due to the highly anisotropic and relativistic jet component, rather than to the accretion disk (see Section \ref{sec:intro}).  The relativistic aberration magnifies the variation amplitudes and foreshortens the time-scales, so that small (10-20\%) monthly or yearly flux changes, that are typical in radio-quiet quasars, become inter- or intra-day variations with amplitudes of a factor of up to 10 or more.  Owing to the distance and to the extreme compactness of the regions at the base of the jet, where most of the radiation is emitted (the light-crossing times inferred from TeV observations are less than an hour), direct imaging of the nuclear zones,  aimed at mapping the structure of the jet and the plasma outflow is not possible.  Rapid variability and correlated variations of the emission at different wavelengths represent  the only effective investigation tool of the jet geometry and dynamics.  Ideally, these should be coupled with high angular resolution radio observations (VLBI) in order to correlate the flux variations at optical, X- and gamma-rays with  the appearance and motion of new radio components moving across and away from the nucleus.   

Multi-wavelength energy distributions and their variability provide some fundamental physical  parameters, most of which are not model-dependent and can thus be considered solid, although affected by the measurement uncertainties and limitations.  As an example, the cooling phase of an outburst (flux decrease) yields the average magnetic field, and the  break frequencies of the synchrotron and inverse Compton spectral components give the electron energy.    Their values constrain the jet models.
Therefore, multi-wavelength observing campaigns of BLLs  are organized in conjunction with outbursts, to study the jet behavior in these sources where its phenomenology is most pure and undiluted.  Observations in the optical domain are crucial, because the emission at these wavelengths  is ``pristine", i.e. not scattered or reprocessed, and rich, i.e. not ``photon-starved" as it is often at X- or gamma-rays, where the photons are normally much fewer.

The first optical studies, started in the 1970's, were based on measurements taken on photographic plates and with photomultipliers \citep{weistrop73,vv1976,miller1978a,milgim1978,westerlund1982,silla1988a}.  More recently, CCD photometry
has replaced the previous types of measurements almost entirely.
Since BLLs are very bright (most of those at  $z \ls $0.5 have magnitudes in the range $V \sim 13-15$), accurate studies of variability are  possible even with small telescopes (0.5-1m aperture) and indeed, the advent of robotic optical and infrared facilities, primarily oriented to the 
follow-up and monitoring of rapid transients (e.g. Gamma-Ray Bursts), has greatly benefitted the BLL field.  These small telescopes (e.g., REM, ROTSE, SMARTS) have 
flexible schedules and pointing constraints, therefore they can be  devoted to long looks of variable sources like BLLs \citep{tosti1996,ciprini2003,carini2004,gu2006,dolcini2007,kasten2011,ackermann2012,bonning2012,nesci2013,sandrinelli2014}.

In few cases, optical orbiting telescopes designed for long, uninterrupted monitorings, like Kepler, have been used for BLL variability studies \citep{edelson2013}.   
Photometry from a satellite also presents the advantage of no atmospheric contamination (``seeing"), which may be an issue at small variability amplitudes, where 
atmosphere instability may mimic intrinsic flickering.  In the ultraviolet domain (1200-3000 \AA), an important role for BLL monitoring was played in the past by the HST High Speed Photometer  \citep{dolan2004}  and especially by the IUE satellite, whose flexibility of scheduling and photometric stability made it apt for ultraviolet variability studies \citep{ulrich1984,hancoe85,maraschi1986,edelson1991,urry1993,palcour1994,gg1997,heidt1997,pian1997}.
Currently,  {\it Swift}/UVOT, which couples good flexibility and sensitivity, is better suited for this class of studies than other ultraviolet orbiting facilities \citep{foschini2007,perri2007,pian2007,tramacere2007,raiteri2010,aliu2013,hesscoll2012}.

Often, optical and near-infrared monitorings of BLLs are organized as part of multi-wavelength campaigns, that have an emphasis on the X- and gamma-ray emission, and  for which the information in the optical domain is crucial to reconstruct the broad-band spectrum for comparison with time-dependent modelling.  This  approach has reached a high degree of sophistication, thanks to the formation of consortia of optical observers
\citep[often including groups of amateur astronomers,  ][]{kato2004}, who have made their expertise and  resources available to this particular purpose.  An instance of this is the {\it Fermi} - {\it AGILE} Support Program, organized within the Whole Earth Blazar Telescope\footnote{GASP/WEBT, http://www.oato.inaf.it/blazars/webt/}. While optical observations coordinated with wide multi-wavelength efforts cover generally outbursts  and high activity states of BLLs \citep{raiteri2009}, standalone optical programs may take place independently of the source flux state and thus offer an unbiased view of  the source behavior in different states \citep{villata2002,villata2004a}.

As already mentioned, variations of BLL emission is seen on time-scales from minutes  to many months and years, with amplitudes from a fraction of a magnitude to a factor of 100 or more  \cite[e.g.]{tosti2002,stalin2006, zhang2008,agudo2011,gupta2012,rani2013}.  Modern observing techniques make it possible to correct reliably for spurious (non-astrophysical) causes of flux variations (e.g. atmospheric instabilities), even when these are small 
\citep[micro-variability: $< 0.1$ mag, time-scales of less than one hour,][]{pollock2007}.  The vast majority of observed optical variations at all time-scales are intrinsic, i.e. not related to interstellar scintillation (irrelevant at optical wavelengths) or lensing by intervening sources, the role of which  was proven to be inconclusive \citep{watson1999,recsto2003,giovannini2004,raiteri2007}.

Often, the rapid variability events violate the limits on efficiency of conversion of kinetic energy into luminosity \citep{guilbert1983}, and/or are inconsistent, based on light-crossing times considerations, with independent estimates of the linear sizes of the inner emitting regions 
\citep[e.g. ][]{sandrinelli2014}.  This suggests lower limits on the  Doppler boosting factors of the order of $\delta \sim 10$, similar to those derived based on observed radio brightness temperatures  and pair-production compactness at gamma-rays (see Section~\ref{sec:intro}).

The correlated optical/near-infrared flux and spectral variability in BLLs can be complex, depending on the source state and on its broad-band spectrum \citep{abdo2010b,nesci2013}.   \cite{bonning2012} and \cite{sandrinelli2014}  report that the color-magnitude diagrams exhibit hysteresis cycles  \citep[as seen occasionally in X-rays, e.g. ][]{takahashi1996,fossati2000}. The former authors also found somewhat larger amplitude variability at the optical than near-infrared wavelengths in BLLs, while the latter authors observed the opposite: this divergence is possibly related to the different time-scales probed.  In low flux states, the optical/near-infrared energy distribution may deviate from a power-law shape and show an excess toward the short wavelengths \citep{sandrinelli2014}, suggesting the presence of a thermal component, that is normally better detected in FSRQs.  

The intra-day variations between different bands in BL Lacertae, the namesake of the 
class and a typical low-frequency-peaked object \cite[according to the definition of  ][]{padgio1995}, are correlated without measurable time lags most 
of the time, and suggest that its optical variability properties remarkably resemble the X-ray variability properties of high-frequency-peaked BLLs. The similarities imply a common origin of the variations, plausibly the most energetic tails of the synchrotron emission produced by the 
relativistic electrons in the jets, for both the optical emission of low-frequency-peaked BLLs and the X-ray emission of high-frequency-peaked BLLs  \citep{zhangyh2013}.

Time algorithms have been devised and applied to BLL light curves in search of characteristic time-scales, recurrences and to measure variability 
indices \citep{heidtwagner1998,ciprini2003,villata2004b,bauer2009,zhangbk2013}.  These parameters give in turn insight into the behavior of the jet at different scales and they can be an effective tool to map its geometry, 
especially when coupled with information at other wavelengths.   However, caution must be used in the interpretation of the outputs of correlation and auto-correlation functions and time structure functions in that often their features do not trace necessarily an authentic and intrinsic characteristic time-scale \citep{emchu2010}.
Unlike other fronts of research on BLLs that have progressed  very quickly in the last two decades,  the field of time analysis still presents facets that are not completely understood, so that despite the availability of advanced observing facilities that allow  sophisticated monitorings in excellent conditions,  the results are still inconclusive.

\begin{figure}
\centering
\includegraphics[width=0.9\textwidth]{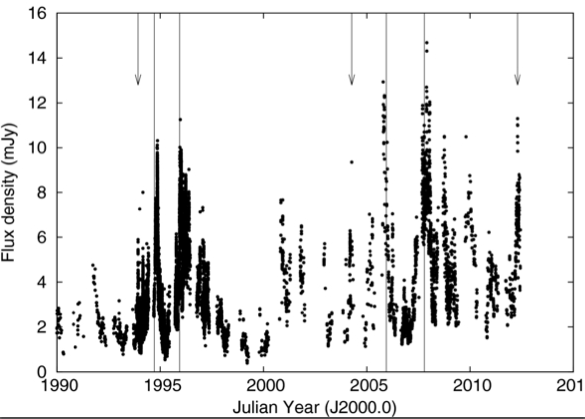}
\caption{Flux variability in the V band of OJ 287 over 25 years.  Note the presence of a number of characteristic recurrent outbursts with an amplitude of an order of magnitude.  The observed variability behavior was suggestive of some periodicity.  The vertical lines represent the outburst timings from the model proposed by \cite{pihajoki2013b}.}
\label{oj287lc}     
\end{figure}

Periodicity has been searched for in the light curves of BLLs at all wavelengths on a wide range of frequencies and never been detected in a consistent way   \citep{rieger2004,osone2006}.  The only exception is OJ~287 ($z = 0.306$), for which a regular flux brightening recurrence every $\sim$12 years was reported  \citep{silla1988b,takalo1994} and confirmed by the detection of 9 equally spaced maxima - of which the latest 4 are very well sampled - in the optical light curve so far  \citep[see Fig. \ref{oj287lc}; ][]{sillanpaa1996,hudec2013,pihajoki2013a}.  The period is ascribed to the presence of a binary system of massive black holes at the center of the 
nucleus, that produce outbursts, often accompanied by precursors, when approaching pericenter  \citep{valtonen2008}.   In these
occasions the flux may brighten by more than a factor of $\sim$100 with respect to quiescence.  Shorter-term variations with periodic and quasi-periodic character were also detected, that may be related to the matter dynamics in the close vicinity of the binary black hole system \citep{pihajoki2013b}.  On the other hand, \cite{hudec2013}  suggest maxima at times that are not foreseen by the model, making the interpretative picture not completely satisfactory.


\section{Polarimetry}
\label{sec:polarim}

Blazars are the astrophysical sources with the highest percentage of  optical linear polarization
in the radio-to-ultraviolet range\footnote{The only other known sources showing similar polarization are
optical counterparts of GRBs, at least during the first seconds to minutes after explosion, and indeed have in common with blazars the non-thermal origin of their continuum.}.
Unlike radio polarization,  the optical polarized signal is immune from propagation effects like Faraday rotation of the polarization plane; moreover, optical polarimetry probes the central nuclear regions of blazar jets,  where the radio emission  is often self-absorbed.  

At optical and near-infrared wavelengths, the  linear polarization percentage can be as high as 40\% and it is normally detected at the level of 5-10\% \citep{angelstock1980,takalo1992,vw1998,tommasi2001a,tommasi2001b,fujiwara2012,pavlidou2013}.
Since blazar optical radiation is  dominated by the synchrotron mechanism, this implies the presence of highly ordered large scale magnetic fields  \citep{westfold1959}. 
Circular polarization could be expected with a degree proportional to $1/\gamma$, where  $\gamma$ indicates the individual  Lorentz factor of an emitting relativistic particle.  Because of this it is very difficult to probe and detect  \citep{takalosill1993}.

High linear polarization  was very soon recognized as a  distinctive feature of blazars, to the point that it has been adopted as a  blazar selection criterion \citep{imptap1988,ilt1991,smith2007,hutsemekers2010,heidtnilsson2011}.   Low-frequency-peaked BLLs and FSRQs are indistinguishable with regard to linear polarization and variability thereof  \citep{vw1998}, while high-frequency-peaked BLLs are somewhat less polarized.  This partially confirms the pioneering study  of \cite{angelstock1980}.

Since interstellar dust can  spuriously polarize the optical light, the detection of  linear  optical polarization percentage must be accurately corrected for this effect by estimating the amount of intervening dust.  On the other hand, the host galaxy stellar light may dilute the intrinsic polarization and lower it artificially, an effect that should also be corrected for \citep{andruchov2008}. However, variability of the polarized signal  -- which is often observed in BLLs -- is a proof of its  intrinsic origin.   

\begin{figure}
\centering
\includegraphics[bb=10 0 490 380, width=0.45\columnwidth]{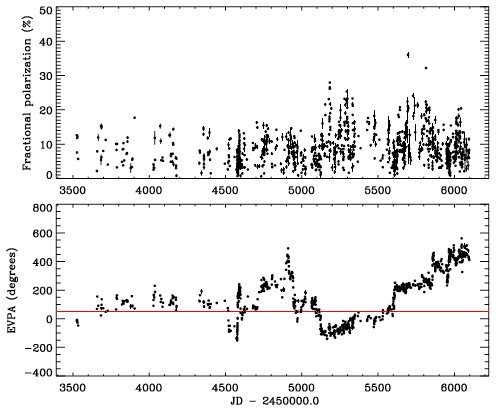} 
\includegraphics[bb=0  50  420 450  ,width=0.43\columnwidth]{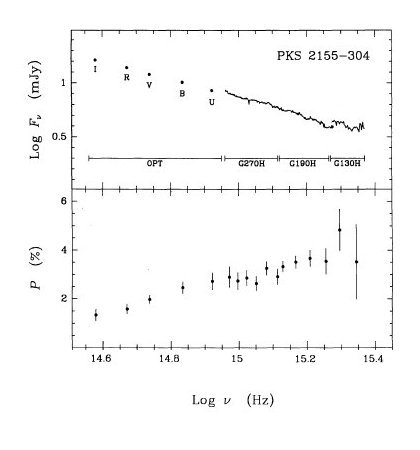} 
\caption{Left:  degree of polarization  (top) and position angle (bottom) of the S5~0716+71 in October 2011  \cite[adapted from ][]{larionov2013}.
Right:  HST and ground-based observations of the flux (top) and polarized degree (bottom) of PKS 2155-304 as a function of  frequency.  The fraction of polarization increases monotonically from optical to ultraviolet by a factor of $\sim$2.
\citep[adapted  from ][]{allen1993}.}
\label{figpolarim}     
\end{figure}

Observations of optical polarization of BLLs have clearly shown that the degree of polarization varies on time-scales from intra-day to years (Fig. \ref{figpolarim}, left), with amplitudes from 20\% to a factor of  2  \citep{imptap1988,tommasi2001a,tommasi2001b,hagenthorn2008}.  In some cases, even faster variations were claimed \citep{impey2000,sasada2008}.  This   variability is consistent with  relativistically beamed synchrotron emission viewed at a very small angle to the line of sight.

Although the polarized optical light variations of BLLs are  occasionally well correlated with those of the total light \citep{sorcia2013}, in general lack of correlation or anti-correlation is observed \citep{gaur2014}.   \cite{barres2010}
proposed that while the optical flux originates in the weakly polarized, stable jet component, the photopolarimetric variability results from the development and propagation of a shock in the jet.  As a  consequence, the optical polarized emission is potentially a better tracer of the high-energy emission, revealing the importance of optical polarimetric monitoring in multi-wavelength campaigns.     Among these, some of the richest  from the point of view of polarimetric coverage were presented by   \cite{urry1997},  \cite{tosti1998}, \cite{pietila1999}, \cite{marscher2008}, \cite{agudo2011}, \cite{larionov2013}, \cite{itoh2013}, \cite{raiteri2013}.  These campaigns were often coordinated with VLBI monitoring of the radio components, and it was noted that the polarization variability events were often correlated with the crossing of the central core by a radio-emitting blob.  In particular, the optical polarization angle settles on the direction parallel to newly born radio knot.
The observed multi-wavelength behavior of the outbursts can be explained within the framework of a shock wave propagating along a helical path in the blazar's jet \citep{camenkrock1992,larionov2010,zhangh2014}.  

In few cases spectropolarimetry of bright BLLs was obtained in optical and ultraviolet and suggested a monotonic increase of polarization percentage towards shorter wavelengths (Fig.~\ref{figpolarim}, right).  From these  measurements  there is no evidence for the presence of  an accretion disk or other thermal source contributing significantly to its  ultraviolet continuum  \citep{allen1993,smith1993}.


\section{Spectroscopy}
\label{sec:specscopy}

\begin{figure}[h]
\includegraphics[bb=  30 50  1500  980,width=0.9\columnwidth]{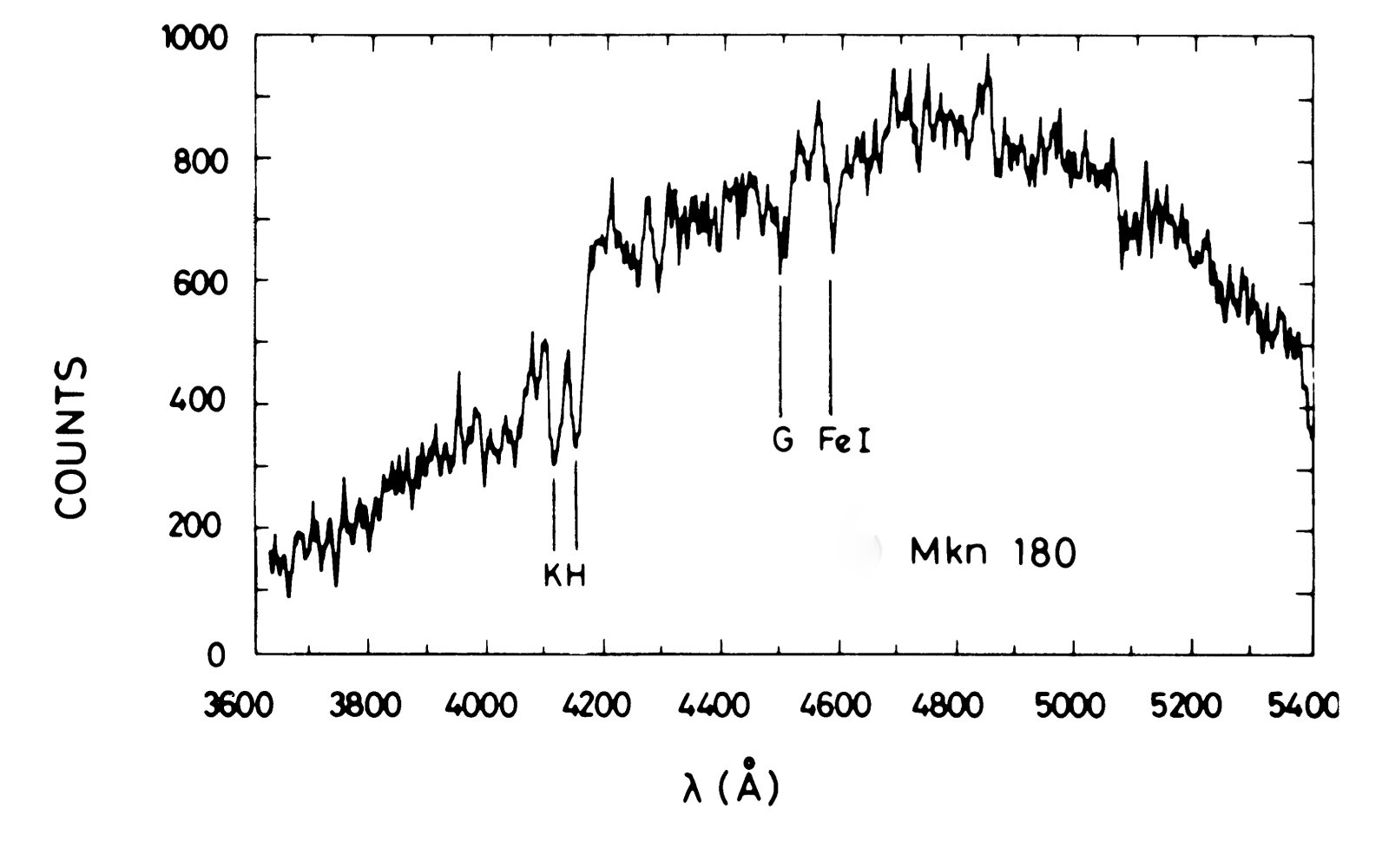}
\caption{The optical spectrum of the BL Lac object Markarian~180 obtained with the first generation of digital detectors.
Absorption features from the host galaxy at $z = 0.046$ are clearly detected \citep{ulrich78}.}
\label{fig:spec_mkn180}      
\end{figure}

\begin{figure}
\includegraphics[width=0.9\columnwidth]{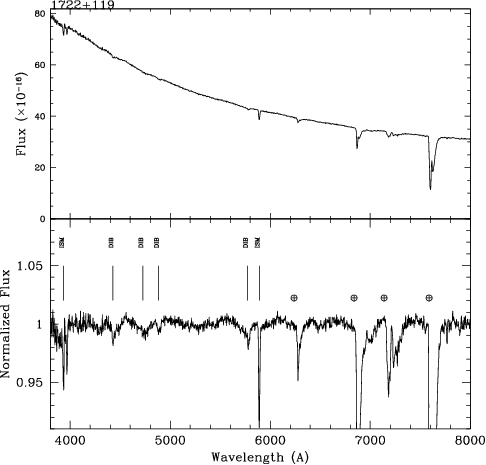}
\caption{The flux calibrated (top panel) and continuum-normalized (bottom panel)  optical spectra of  H~1722+11. The only absorption features  besides the telluric bands are due to the Galactic interstellar medium:
Ca II  $\lambda\lambda 3934, 3968$ and Na I $ \lambda5892$ atomic lines and  DIBs at 4428, 4726, 4882, and 5772 \AA 
\citep{sbarufatti06b}. See http://archive.oapd.inaf.it/zbllac/ for more examples.}
\label{fig:spec_1722}      
\end{figure}

The (quasi) featureless optical spectra of BLLs are historically the main characteristic 
for this class of objects, that distinguishes them within the blazar family (see Section \ref{sec:intro}). 
This peculiarity made them rather elusive, because of the consequent  
difficulty to determine their redshift and thus the distance. 
However, it was early recognized that for a number of low redshift BLLs it was possible to 
detect faint absorption  and/or emission  features
\cite[see e.g., ][and also Section \ref{sec:hostgals}]{miller1978a,miller1978b} 
that allowed one to prove the extragalactic nature of the sources and assess their redshift.
For low redshift targets the optical spectra  (see example in Fig. \ref{fig:spec_mkn180})
left no doubts that the BLL phenomenon occurs in the 
nuclei of galaxies dominated by old stellar populations \citep{ulrich78}, a result that was also confirmed 
by first imaging studies (see  Section \ref{sec:hostgals} for details). 
In most cases, however, the high luminosity of the non-thermal emission from the nucleus with respect to the light from the 
underlying nebulosity  yields spectra with very weak lines. When observed with poor signal-to-noise ratio and resolution 
their optical spectra appear therefore featureless  (see example in Fig. \ref{fig:spec_1722}).

The spectral features of BLLs can be grouped into three types: 
1) spectral lines of stars from the host galaxy;
2) weak emission lines characteristic of low density gas; 
3) intervening absorptions from cold gas.
While from the first two types a redshift  can be derived, 
in the latter case only a lower limit on the redshift can be obtained. 
The faintness of intrinsic absorption or emission lines strongly depends on the 
brightness level of the nuclear source that is dominated  by the jet component. 
Given the significant flux variability of the nuclear source (see Section \ref{sec:opticalvar}), the detection of  features 
in the spectra depends on the brightness state of the BLL. Faint states  favour 
the detection of intrinsic emission lines while during high states intervening absorptions may be more 
easily detected because of the better signal-to-noise ratio.

\subsection{Broad emission lines}
\label{sec:bremlines}

\begin{figure}[h]
\includegraphics[bb= 10 180 540 570, width=0.9\columnwidth ]{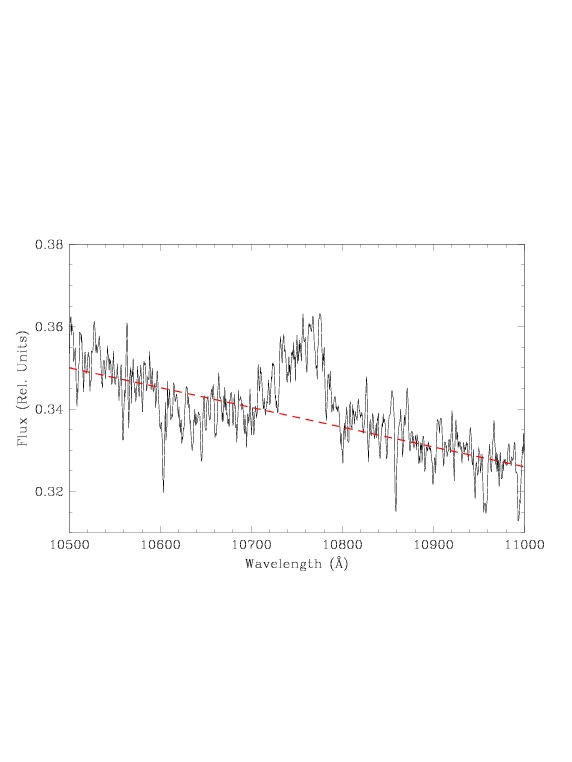}
\caption{The  optical spectrum of PKS 0048-09 showing a very faint (equivalent width $\simeq$ 3 \AA)
broad (2000 km s$^{-1}$) emission line of H$\alpha$ at $z = 0.638$.  The power-law spectral continuum is shown as a red dashed curve
\citep{landoni12}.  See http://archive.oapd.inaf.it/zbllac/ for more examples.}
\label{fig:spec_0048}      
\end{figure}

Although the  optical  spectrum of BLLs is characterized by a featureless continuum, in a number of cases (see example in Fig. \ref{fig:spec_0048}) 
weak broad emission lines, similar  to those observed in quasars but at lower intrinsic luminosities, are seen.   Owing to the weakness of these lines they may be detectable depending on the state of the source  and the quality of the observations.  Such broad emission lines were occasionally found also in the prototype of the class, BL Lacertae  \citep{vermeulen1995,corbett1996}.\footnote{As an aside, the detection of a broad  H$\alpha$ emission line at the redshift of the host galaxy of BL Lacertae proved that the BLL prototype is not a high-redshift micro-lensed quasar \citep{ostriker1985,ostriker1990}.}
Comparing the spectra of BLLs and FSRQs it was found  indeed that there is a continuity of
emission line properties between the two sub-classes of blazars  \citep{scarpa97}, depending on the relative strength of
the various involved components (jet emission, thermal disk or broad line region emission, host galaxy). 
It turns out therefore that the  classification of an object as BLL or FSRQ may depend on the spectroscopic adopted criteria \cite[see e.g.][for a detailed discussion on this issue]{padovani2012}.  
The emission line photons may be comptonized by both cold and 
relativistic electrons in the jet \citep[external Compton, ][]{sikora1994}.  While in FSRQs, owing to their conspicuous broad emission 
line luminosity, this is the dominant mechanism for production of gamma-rays, in BLLs it never exceeds significantly self-Comptonization of 
synchrotron radiation.

\subsection{Host galaxy lines}
\label{sec:hostlines}

\begin{figure}
\includegraphics[ width=1.0\columnwidth ]{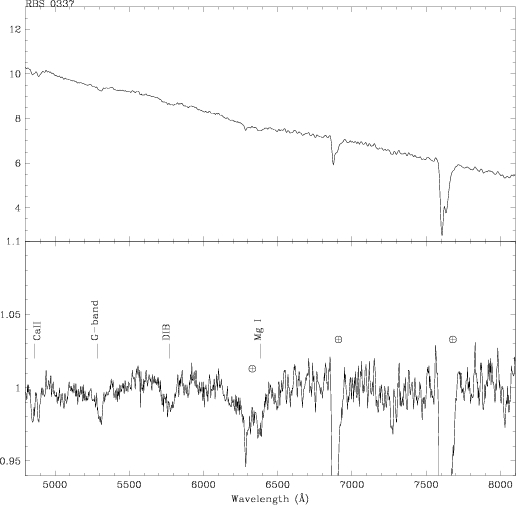}
\caption{The  optical spectrum of  RBS 0337 obtained at VLT (\cite{landoni12}.
Superposed on the bright non-thermal emission are a number 
of faint absorption lines from the stellar population of the host galaxy.  Telluric and interstellar medium features are also marked.
The bottom panel shows the normalized spectrum.
See http://archive.oapd.inaf.it/zbllac/ for more examples.}
\label{fig:spec_0337}      
\end{figure}

As discussed in Section \ref{sec:hostgals} virtually all 
BLLs reside in massive elliptical galaxies (or galaxies with a prominente spheroidal component).
This means that their stellar population is dominated by old stars and the main absorption  features 
are Ca II 3934,3967 , G band 4304, H$_\beta$ 4861, Mg I 5175, NaI 5875. These lines can be detected 
over the non-thermal component depending on the relative power of the two components (non-thermal and host galaxy) 
and on the signal-to-noise ratio and spectral resolution of the observations. 
Early spectroscopic studies of BLLs \citep[see e.g., ][]{ulrich78} were limited by relatively 
modest spectroscopic capabilities and many objects remained therefore classified as featureless. 
Only for objects at low redshift and with faint nuclei these absorption lines 
could be observed since the bright continuum swamps the thermal component thus reducing significantly their equivalent width.
High quality spectroscopic observations  obtained with  large  telescopes \citep{sbarufatti05,sbarufatti06a,landoni12,sandrinelli2013,shaw2013} improved  significantly the ability to discover  faint features from which the redshift is  derived (see example in Fig. \ref{fig:spec_0337}) 

In very few cases high quality spectroscopy secured with high spatial resolution probed 
ongoing star formation in the inner  region of the host galaxy through the detection of narrow emission lines.
A noticeable case is PKS~2005-489 ($z = 0.071$) for which 
spatially resolved emissions were observed  in an extended rotating ring, 
perpendicular to the position of the radio jet, at 4 kpc from the nucleus. 
The presence of this rotating star-forming region is the signature of a minor accretion event that has 
funneled gas toward the central few kiloparsec region around 
the active nucleus \citep{bressan2006}.

%
\subsection{Intervening features}
\label{sec:intlines}

As in the case of high redshift quasars,  gas along the line of sight to BLLs  may produce 
absorption systems at a number  redshifts lower than the redshift of the object.
Since BLLs  exhibit very weak or no intrinsic absorption lines,  they are ideal
sources for studying  intervening absorptions. In particular they  may be due to the Galactic
interstellar medium. In the optical the Ca II and Na I are noticeable, together
with  diffuse interstellar bands (DIB) of molecular origin.  An example is given in  Figure \ref{fig:spec_1722}.
   

\begin{figure}[h!]
\centering
\includegraphics[width=0.7\linewidth]{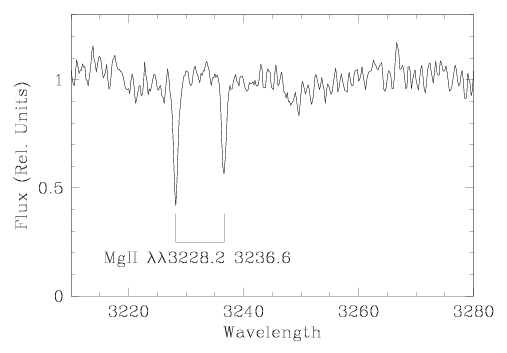}
\caption{ESO VLT X-Shooter optical spectrum of PKS 0048-09 (z = 0.635) showing  intervening Mg II absorption  doublet at $z = 0.154$ \citep{landoni12}
}
\label{Sbarufatti2006}
\end{figure}
   

The bulk of baryonic matter ($\sim90\ \%$) probably resides around and between  galaxies at low redshifts.   
A large fraction of it  ($\sim40\ \%$) is at high temperature  ($T>10^6\  K$).  It has been probed
through the observation of O VII and O VIII and other X-ray absorption lines. In particular we recall here the observations of 
 the  bright BLL PKS 2155$-$304 \citep[e.g.][]{nicastro2002,cagnoni2004}.  
At lower temperatures ($<10^4\  K$) the intervening gas can be traced by the Ly$\alpha$ forest absorption, while absorptions 
associated with galaxies halos are detected from metals (mainly Mg II and C IV). 
Relevant results in connection with BLLs and this line of research are consequent to the installation into HST of the Cosmic Origin
Spectrograph (COS), which is a medium-high ($R \sim 10000$) resolution instrument of
unprecedented sensitivity  \citep[e.g.][]{green2012}. 

\begin{figure}[h!]
\centering
\includegraphics[ width=1.0\columnwidth ]{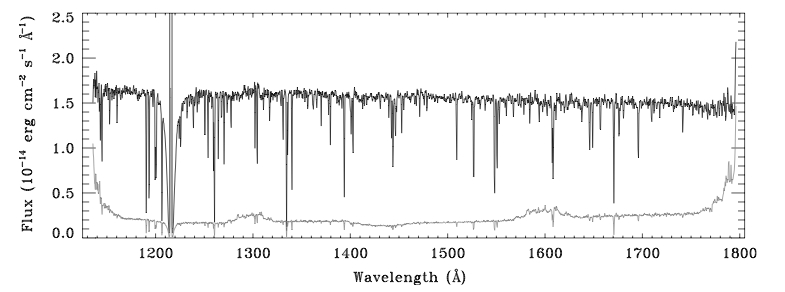}
\caption{Ultraviolet COS spectrum of PG 1553+113. Several narrow Lyman  absorption line systems at different redshifts are observed \citep{danforth2010}. A lower limit on the redshift of the BLL  can be obtained from the distribution of the redshifts of the absorption systems.
}
\label{Danforth2010}
\end{figure}

So far, spectra of   four  BLLs (PG~1553+113, S5~0716+714, 3C~66A, and PG~1424+240)  of previously unknown redshift  were obtained 
 in the wavelength range 1135-1795 \AA.  The data show a
smooth continuum, with narrow ($\sim$100 \ km s$^{-1}$) absorption features
arising mainly from the Lyman series. The redshift of each Ly$\alpha$
absorber can be measured in the  $0 \lesssim z \lesssim 0.43$ interval .
The maximum redshift  yields immediately a lower limit to the redshift of
the BLL, while an upper limit is derived from  the expected distribution of the
absorbers.    An example of a spectrum is given in Figure \ref{Danforth2010}.  An
important aspect of these findings is that all sources are GeV emitters
(three of them are also TeV emitters). At TeV energies and $z\sim0.4$, the opacity of
the extragalactic background light for pair production is
$ >1$  (see Section \ref{sec:mwlref}).  Using this technique a TeV source (PG 1424+240) 
was found at z $>$ 0.6.
The redshifts derived from  COS data, together with the gamma-ray
spectra  are therefore an effective  powerful probe of the extragalactic infrared background  \cite[e.g.][and
references therein]{costamante2013}. 
   
A similar technique can be applied using  Mg II  $\lambda2800$ absorption doublet, which at
$ z> 0.15$ can be observed  from the ground.  In Figure \ref{Sbarufatti2006} we report the  ESO VLT X-Shooter optical spectrum of PKS 0048-09 where a Mg II absorption at $z = 0.154$ is
apparent. 

\section{Host galaxies}
\label{sec:hostgals}

\begin{figure}
\includegraphics[width=0.49\columnwidth, bb =0 0 400 400]{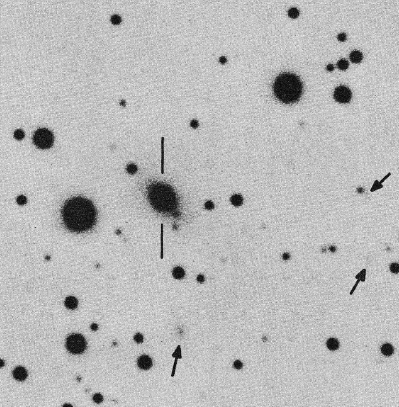}
\includegraphics[width=0.5\columnwidth, bb =0 0 415 415]{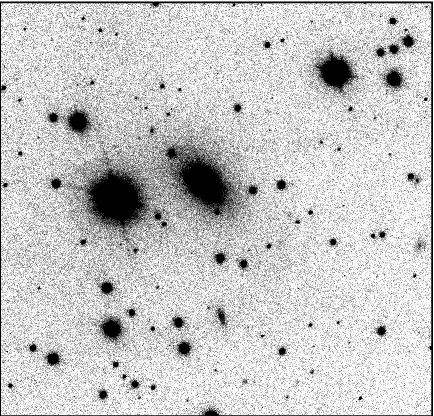}
\caption{Left:  BL Lacertae as imaged in a photographic plate (Kodak IIa-J) 
in 1974 at the prime focus of  Kitt Peak 4m telescope \citep{kinman1975}. 
The target, indicated by two vertical lines, is clearly elongated and surrounded by a 
faint nebulosity.   Arrows mark faint extended sources.  Right: The same field as imaged in the R filter and CCD detector 
by the 2.5m Nordic Optical Telescope in 1998.}
\label{fig:bllac}      
\end{figure}

Soon after their discovery, it was noted that BLLs could be surrounded by a faint luminosity  \citep{disney74a,disney74b,stein76,miller1977}. 
In the late 1970s the use of modern detectors (CCD) allowed observers to probe with better accuracy the nature of the nebulosity (e.g. Fig. \ref{fig:bllac}).  \cite{weistrop79} imaged the BLL PKS 0548-322  in various filters and found it to be composed by a giant elliptical galaxy (M$_V \sim $ --22) with a bright nucleus. About 10 years later, at the second conference entirely dedicated to BLLs in Como, M-H. Ulrich  \citep{ulrich89} reported  about 15 objects for which a host galaxy was
measured and compared their  properties with those of bright radio galaxies in order to test the consistency of the
hypothesis that BLLs are indeed the parent population of FR - I radio galaxies with the jet pointing closely to  the observer direction. 
With the caveat of the exiguity and non-homogeneity of the  dataset it was
argued that the BLL host population is  very similar to that of radiogalaxies.

The following decade has seen a significant improvement in terms of quality and sample size 
of BLL host galaxies.  Various authors \citep{abraham91,stickel93,falomo1996,wurtz96,falomo1999} 
undertook ground based optical imaging studies of sizeable samples of objects 
including radio and X-ray-selected objects, according to the old definition. 
All these observations led to the conclusion that BLLs are the active nuclei of  giant/massive elliptical galaxies with 
average luminosity in the R band M$_R$ $\sim$ -- 23. 
Only very few cases were reported to be different from this scheme suggesting that 
the host is a disk dominated galaxy \cite[example: 1415+255; 1413+135;][]{stocke92,halpern86} but these cases 
were rather controversial and mostly due to the modest angular resolution of the images \cite[see e.g.][]{gladders97,falomo1997a,lamer99}.

Optical observations were also complemented by near-infrared images, that would favour the 
detection of the host galaxy with
respect to the nuclear emission owing to the difference of the ratio of the two components. This is more relevant for
relatively high resdshift sources since the emission  from the host galaxy peaks in the near-infrared while in the optical
band the light from the underlying stellar population diminishes significantly for objects with z $>$ 0.4-0.5. Near-infrared
observations of BLLs were collected by \cite{wright98} and \cite{kotilainen98} who extended the sample of resolved objects
confirming previous results derived from optical studies. Moreover in a number of cases both optical and near-infrared
images were available thus allowing the estimate of the integrated broad band color (R-H) on BLLs hosts. It turned out
that the color is indistinguishable from that of inactive elliptical galaxies of similar luminosity
\citep{kotilainen98}.

In  spite of these imaging studies  a significant fraction of observed objects remained 
unresolved and often these targets are also of
unknown redshift. This is because, in order to disentangle the light of the 
host galaxy from that of the bright nucleus it is
mandatory to observe the nuclei with the best possible angular resolution (i.e. narrow point spread function)
and also have adequate signal-to-noise ratio in order to characterise the faint extended emission around the nucleus.

\begin{figure*}
\includegraphics[ angle=-90,width=1.0\columnwidth]{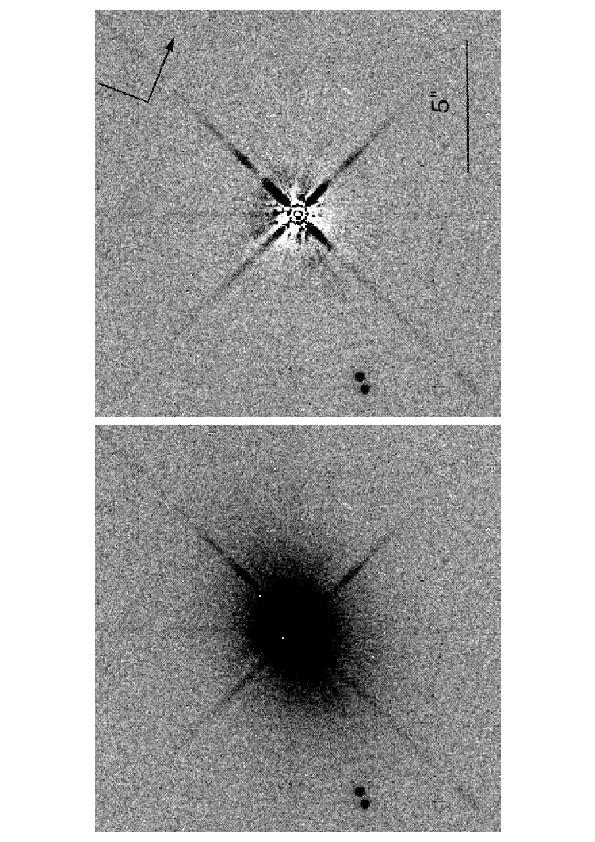}
\caption{Image of BL Lacertae obtained by HST (left) compared to galaxy-model subtracted image (right). No significant residuals are detected after the subtraction of the model (point source plus elliptical galaxy, convolved with the point spread function).
The cross-shaped residuals are due to the diffraction
pattern of the HST point spread function, which is not fully described by the two dimensional
modelling \citep{falomo2000}. 
}
\label{fig:hstima}      
\end{figure*}

This achievement became possible with the observations of the refurbished HST and WFPC2. The first images of BLLs \citep{falomo1997b,jannuzi97,urry1999} clearly demonstrated the great advantage of the HST angular resolution  in  characterizing the properties of the galaxies (see example in Fig. \ref{fig:hstima}). 
The success of the first images prompted the acquisition of a large dataset of 
images through an HST snapshot program. Using
the WFPC2 in the F702W filter short exposure (300-1000 s) images of 110 BLLs were secured \citep{scarpa00}.
These form the largest and most homogeneous dataset of BLLs high resolution images. 
In this dataset it was possible to
resolve all BLLs at redshift $<$ 0.5 and many others at higher redshift or with unknown z. All resolved host galaxies
were found to be well fitted by an elliptical model (or with a dominant bulge component) and exhibited a relatively narrow
range of absolute magnitude \cite[M$_R$ = --23.0 $\pm$ 0.6;][]{urry2000} 
with no significant difference depending on the original 
selection procedure (radio or X-ray). Comparison with the properties of nearby radiogalaxies strongly supports the
unification picture with FR I galaxies as a favorite  parent population \citep{urry95}.

\begin{figure}[h]
\includegraphics[bb= 130 300 465 750 ,width=0.42\columnwidth ]{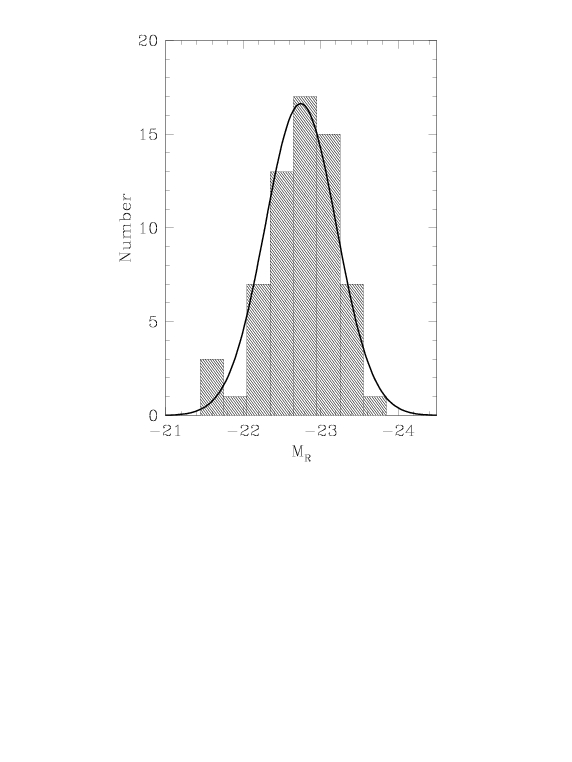}
\includegraphics[bb= 0 180 530 680 ,width=0.6\columnwidth ]{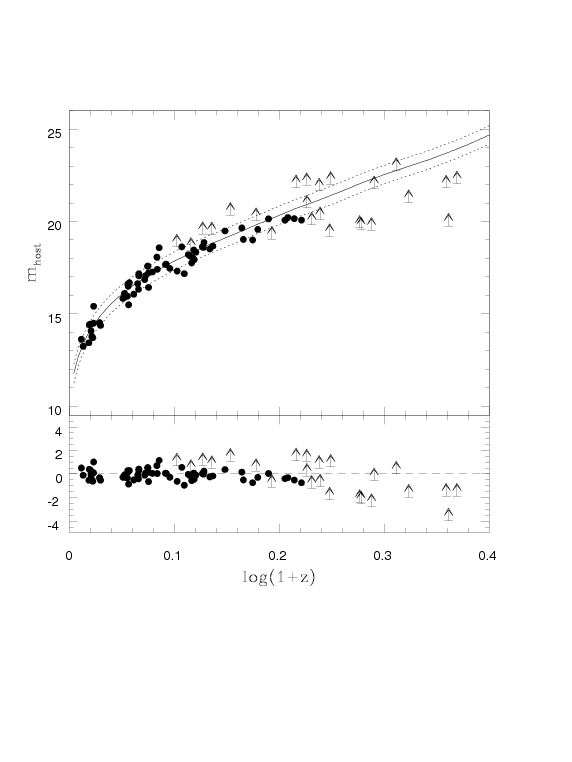}
\caption{Left: Distribution of the host galaxy absolute magnitudes M$_R$ of BLLs from HST observations \citep{sbarufatti05}.  The solid line represents a Gaussian fit to the distribution (mean M$_R$ = --22.9). Right: Hubble diagram for host galaxies of BLLs (filled circles).  Magnitude limits on unresolved host galaxies   are indicated  with arrows. The solid
line corresponds to  a galaxy of  M$_R$=--22.9. Dotted curves show a  0.5 magnitudes spread.  The scatter of the observed data with respect to the solid line is shown in the lower panel \cite[see ][for details]{sbarufatti05}.}
\label{fig:hstmabs}      
\end{figure}

A detailed morphological analysis of the nearest (z $<$ 0.2 ) objects \citep{falomo2000} 
 revealed that the host galaxies
 are mostly smooth and unperturbed ellipticals with no signature of dusty features or
 sub-structures, albeit with a high incidence  of close companions (see Fig. \ref{fig:hstima}). 
 It suggests that strong gravitational interaction is not relevant for the
 ongoing activity. Moreover in all cases the nucleus was found very well centered on the main body of the galaxy arguing
 against a microlensing hypothesis for the interpretation of the class \citep{ostriker1985} .

The characterisation of the host galaxies could also help to constrain the distance of the objects in the cases 
of pure featureless optical spectra.
To this aim it is mandatory to use  high quality images  of the targets. From the
whole ensemble of imaging studies of BLLs, it turned out that the range of absolute magnitudes of the host
galaxies of this class of AGN is relatively narrow and therefore one can use their 
luminosity as a sort of {\it standard candle}. Under this assumption (see Fig. \ref{fig:hstmabs} ) if the host is 
detected then one can derive the {\it imaging redshift} from the apparent magnitude of the host galaxy. The method, initially
proposed and applied to ground based images \citep{romanishin87,falomo1996,falomo1999} was then expanded and perfected  using HST images \citep{sbarufatti05}.  Providing that the photometric and structural properties of the host galaxies are well determined (by high quality images) the method is relatively robust and can yield redshift estimate with an average accuracy of $\Delta z \sim$ 0.05   \cite[see Figure 4 in][]{sbarufatti05}.
Also in the case of non detection of the host galaxy it is
possible to set constraints on the minimum redshift of the object  \cite[see e.g. ][]{treves07}.

\section{Jets}
\label{sec:mwljets}

Radio-loud AGNs  are often associated with jets.  The quasar prototype 3C 273 was very early recognized to exhibit an optical jet structure \citep{schmidt1963}. In the case of BLLs this association is  even more cogent, since they are characterised by relativistic jets pointing in the observing direction  \cite[see ][and Section \ref{sec:intro}]{blarees1978}. 
Nevertheless,  the very fact of being pointed to the observer may make the study of the jet morphology  rather elusive.
In the radio band, thanks to the high angular resolution, jet detection is possible and in fact a
large fraction of objects classified as BLLs show the signature of a jet, often with evidence of superluminal motion  \cite[see e.g.][]{giroletti2004}.  
The arcsec structures are well studied with VLA or similar instruments \citep{ulvestad86,laurent1993}.
Radio jet sub structures are commonly detected  with milliarsec resolution when observed with the
VLBI  \cite[][and references therein]{ojha2010,piner10}. At z $\sim$ 0.1 one can thus explore  structures down to parsec scale. 

Some counterparts are found also in X-rays in Chandra images (angular resolution $\sim$1 arcsec) at relatively high redshift. 
They are interpreted as a result of the Compton scattering cosmic microwave background photons off the most energetic electrons \citep{schwartz2000,tavecchio00,celotti01}. 
Jet morphology in X-ray was studied in some details for a number of close-by BLLs, like  PKS~0521--365 
\citep{birkinshaw2002}, and PKS 2201+044 and 3C 371 \citep{sambruna2007}, 
which notably are systems where the beaming is supposed to
be modest. Knot structures are  observed.

\begin{figure}[h]
\includegraphics[bb= 10 10 580 440, width=0.95\columnwidth]{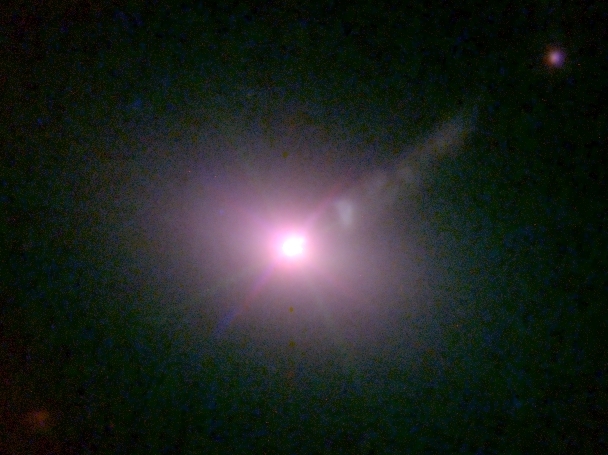}
\caption{Combined color image (optical by HST + WFPC2; near-infrared by VLT + MAD) 
of the jet of PKS 0521--365.  The jet is clearly structured  with the closest  knot at  $\sim 1.5^{\prime}{\prime}$ from the bright nucleus. The inner part of the giant elliptical galaxy that hosts this BLL is also apparent. The faint red  tip emission on the top right is not associated with the jet \cite[see ][for details]{falomo2009}.}
\label{fig:imajet0521}      
\end{figure}

On the other hand, the jet is  rarely detected in the optical and near-infrared bands. 
This depends both on the more limited angular resolution with respect to the radio band and on the short lifetime of the high energy
electrons  producing the non-thermal emission of the jet. 
Because of the requirements on spatial resolution, in the optical and infrared the jet morphology was
mainly studied with HST \citep{scarpa99} and more recently with  adaptive optics methods \citep{falomo2009}.
A detailed study of the optical jet based on HST images was completed for 
PKS 0521--365, PKS 2201+04 and 3C 371 \citep{scarpa99}. 
In Figure \ref{fig:imajet0521} we show an example of optical jet of the BLL PKS 0521--365 
(z = 0.0554) superposed onto the prominent massive host galaxy.


\begin{figure*}
\includegraphics[width=1.0\columnwidth]{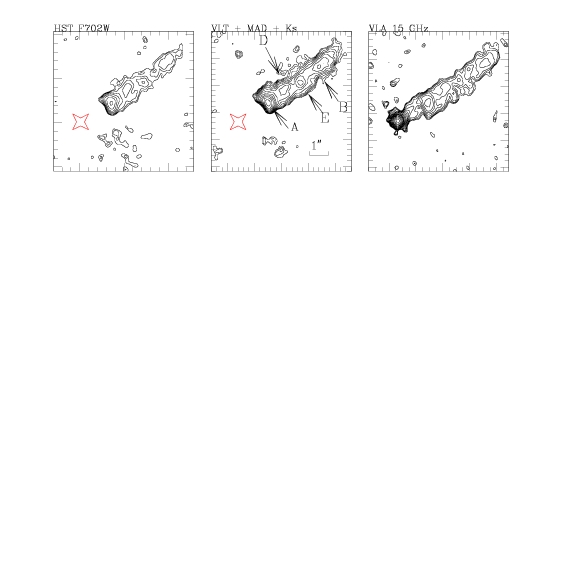}
\vspace{-7cm}
\caption{The contour plot of the jet 
of PKS 0521--365 observed by HST+WFPC2 in R band (left),  VLT + MAD in the $K_s$ band (middle) and VLA at 15 GHz (right). The large star represents the position of the (subtracted) nucleus in both
optical and near-infrared band \citep{falomo2009}.}
\label{fig:jet0521}      
\end{figure*}

Detailed analysis of the jet structure at radio, near-infrared and optical frequencies reveals 
a number of knots (see Fig. \ref{fig:jet0521}) that exhibit remarkable similarity at all frequencies  \citep{falomo2009}.


\section{Environments}
\label{sec:environs}

After the recognition  that BLLs are hosted in elliptical galaxies an obvious further 
step was to search  for associated groups or
clusters of galaxies. Early suggestions of such association in the case of PKS 0548-322 \citep{disney74a} turned out to be false (but see below for surprise) and few other cases were suggested for 3C66A, AP Lib, PKS 1400+164 \citep{butcher76,visvanathan77,weistrop83} but not properly confirmed. For instance in the case of 1440+164, for which an association with a group of galaxies was claimed, deep images obtained with modern instrumentation complemented by
spectroscopy of a number of these galaxies showed that the group is not physically associated  
with  the BLL \citep{pesce94}.
 
\begin{figure}[hb]
\includegraphics[width=0.47\columnwidth ]{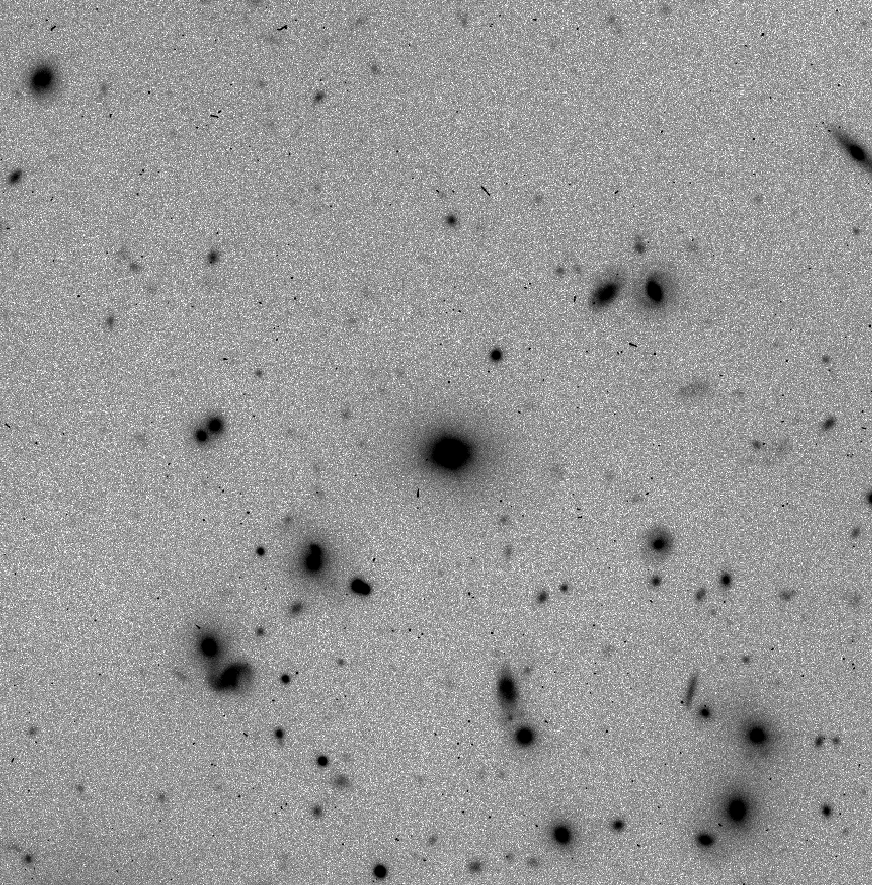}  
\includegraphics[bb= 0 20 1010 1000, width=0.505\columnwidth ]{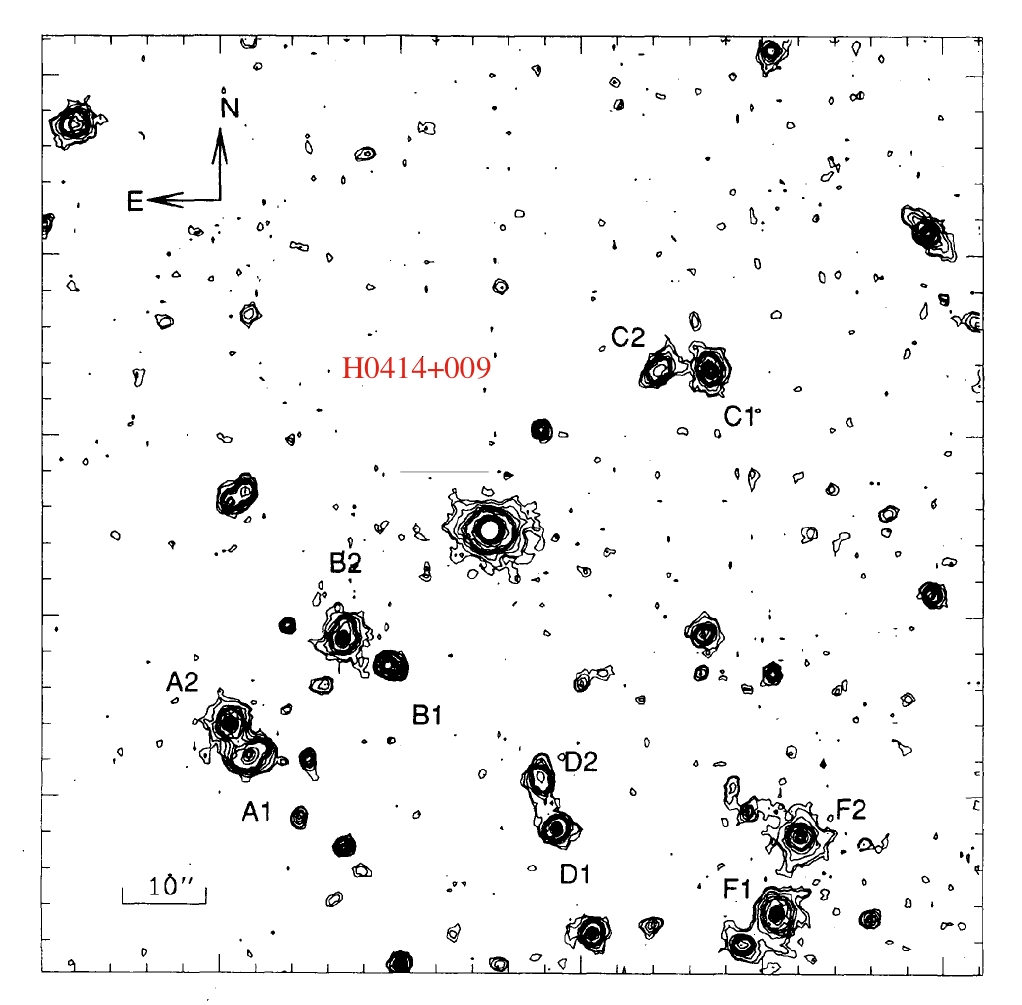}  
 \caption{Left: ESO NTT R-band image of the  galaxies in the field of H 0414+009.    Right: Contour plot of the image, showing galaxies at the same redshift as the BLL, forming a physical group of which the BLL host is the dominant member \citep{falomo1993c}. }
 \label{fig:h0414}      
\end{figure}
 
It was  25 years after the discovery of the BLL class that a number of systematic studies of the cluster environments
around BLLs were undertaken. Using optical images \cite{fried93} performed a study of the galaxy density for the 1 Jy
sample and found an increased galaxy density for low ($z \sim 0.3$) and intermediate ($z \sim 0.6$) redshift sources but
not for higher redshift objects ($z \sim 1$). However, for the latter subsample their images were not deep enough to probe
adequately the environments. Subsequent studies \citep{wurtz93,smith95,wurtz97}  showed that on average BLLs inhabit poor galaxy environments (Abell richness class $\sim$ 0). Attempts to compare the environment properties with those of the alleged parent population (see Section \ref{sec:intro}) of FR I radiogalaxies yielded somewhat controversial results
\citep{smith95,wurtz97}.   All these studies were based only on photometry of the fields around the BLL sources,
therefore the association with group of galaxies can be assessed only basing on statistical 
considerations. For a number of selected targets these imaging studies were complemented by spectroscopy of galaxies in the fields that demonstrated the physical association (same redshift) between the galaxies and the BLL source  \cite[see e.g.][]{falomo1993b,falomo1993c,pesce94,pesce95}.


\begin{figure}[h]
\includegraphics[width=0.95\columnwidth ]{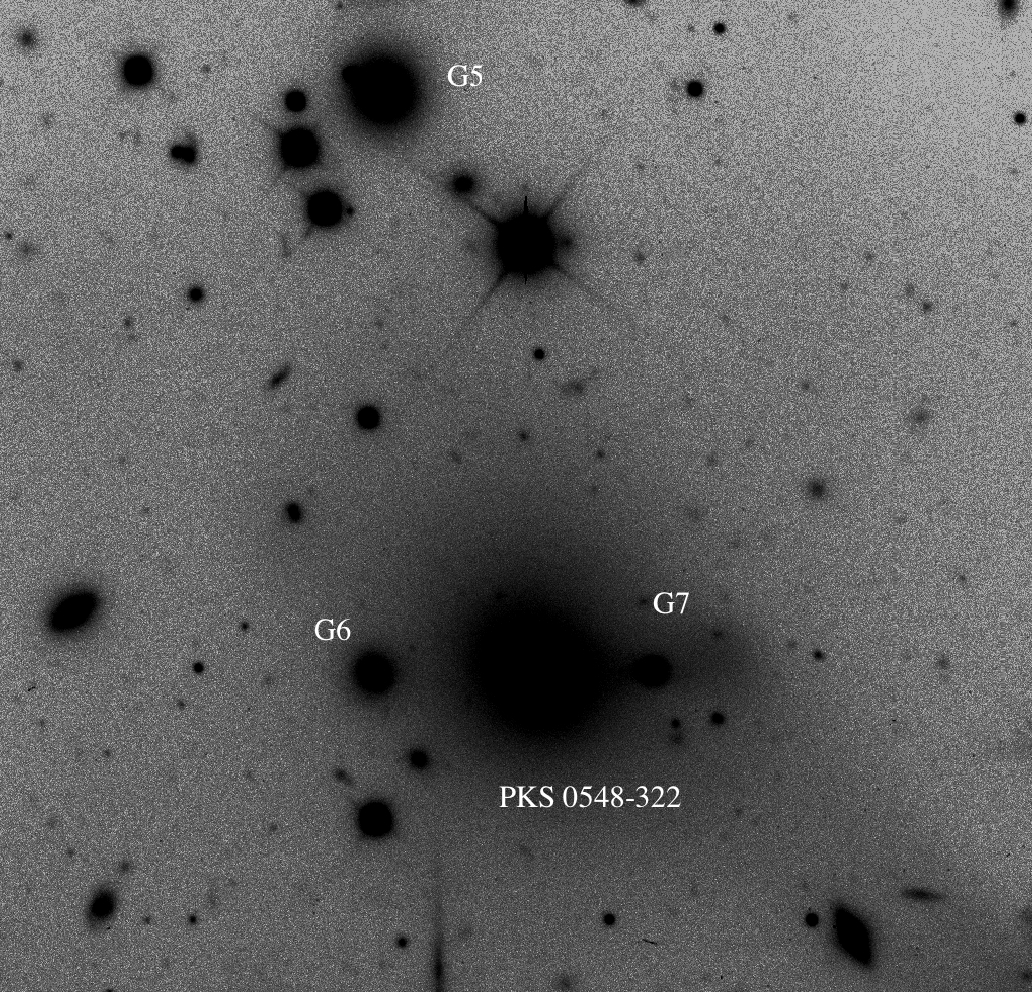}
\caption{The field around PKS 0548-441 imaged at ESO NTT + SUSI (R filter) showing the host galaxy 
and a number of galaxies at the same redshift as the BLL. This is a rare example of BLL associated with a relatively rich cluster of galaxies. \citep{falomo1995}.
}
\label{fig:p0548}      
\end{figure}

\begin{figure}[h]
\includegraphics[bb=15 110 475 650, width=0.95\columnwidth ]{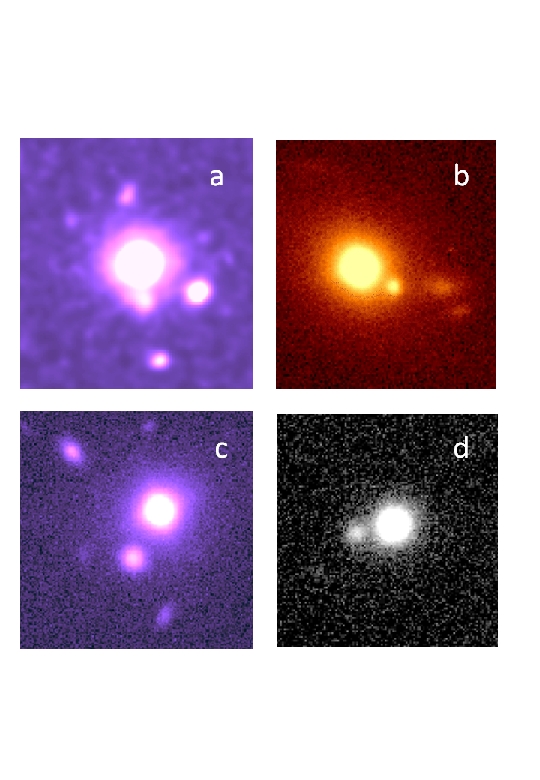}
\caption{Close (few arcsec angular distance) companions of BLLs (central bright objects): (a) PKS~0301--243:  deconvolved image showing companions and a jet-like feature;
(b) H~0323+022: distorted morphology and close companions;  (c) PKS~0829+04:  extended companions;  (d) PKS~2335+031: close resolved companion galaxy \citep{falomo1996}.
}
\label{fig:cc}      
\end{figure}

A  interesting case of rich environment is found for the radio source PKS 0548-441. This target was
among the first BLLs to be pointed out as likely connected with a cluster of galaxies at $z \sim 0.04$ \citep{disney74a}.
The claim was, however, soon disproved as the redshift of the source (z = 0.069)  was derived \citep{fosbury76}. This nearby object was deeply investigated by \cite{falomo1995} who obtained both high quality images and spectroscopy of galaxies in the field . It turned out that the source is indeed located in a rich cluster of galaxies (see Fig. \ref{fig:p0548}) of which the BLL host galaxy is clearly the dominant member with an absolute magnitude M$_R$ = --23.5. The physical association of the cluster was proved by spectroscopy of many galaxies in the field that have the same redshift as the BLL source. This is a rare case of BLL object associated with a relatively rich (Abell class $\sim$ 2) cluster of galaxies while most of the BLLs reside in poorer environments.

In spite of the relative bareness of BLL  galaxy environment,  it is rather frequent to find close
companions within 10 arcsec (projected distance $<$ 20-50 kpc for redshift in the range 0.1--0.4) from
the target. Relevant examples are shown in Figure \ref{fig:cc} \citep{falomo1996}. In some cases the
companion was found to be physically associated with the active nucleus \citep{falomo1993b,falomo1995,heidt99}. 
However, owing  to the faintness of the companions and their proximity to the bright BLL source, spectroscopy of the companions is lacking in many cases. Moreover in some cases also the redshift of the BLL target is unknown, and it is not possible to prove the association.


\section{The black hole mass}
\label{sec:bhmass}

In AGNs the mass  of the black hole ($M_{BH}$)  is a basic parameter characterizing the source. The standard 
paradigm is that AGN are supermassive accreting black holes \citep{salpeter1964,zeldovich1964}.  
The mass fixes an upper limit to the AGN  luminosity through the argument that the Eddington luminosity
cannot be significantly overcome. If the accretion occurs through a disk, its inner radius depends
on the mass, because it is related to the gravitational radius, and in turn this constrains the
internal temperature. The disk inner dimension gives a number of natural time-scales, as the light
crossing time, or the Keplerian time. The mass therefore enters as a basic parameter 
 in the interpretation of the AGN variability.

The  most widely used method to estimate the mass of  quasars or Seyfert galaxies nuclei 
is based on the assumption that broad emission lines are produced by fluorescence of cold clouds
irradiated by a central source, and orbiting the central black hole.  The line width 
is  related to the  velocity of the clouds while the distance of the cloud may be 
derived from  \textit{reverberation mapping}, that  yields the time delay between a flare
in the continuum, and the response of a line flux which is produced in the cloud 
\citep{peterson1997}. Because of the expensiveness of this method, alternatively one can use an  
experimental relationship (with significant dispersion) 
between the size of the broad line region and the luminosity of the continuum close to the emission line.
This technique allowed the estimate of the mass of thousands of quasars  with a  precision of a factor  $\sim$ 3 \citep{shen2011}.

In the case of BLLs this method can be applied only to few sources showing broad emission lines and  with particular caution, 
since the observed emission is dominated by a relativistic jet. 
Therefore it is of obvious importance to adopt other methods to estimate the black hole mass 
of BLLs.  In the local Universe it was demonstrated that most galaxies contain in their center a massive 
black hole and that its mass is correlated with some properties of its host  galaxy 
(luminosity, mass, velocity dispersion).
As a rule of thumb the mass of the black  hole is $\sim$ 0.1\%  of that of the spheroidal component.
The host galaxies of BLLs are supposedly giant ellipticals, i.e the whole galaxy coincides with
spheroidal component. The mass of the ellipticals scales as the luminosity. Therefore one can construct a
black hole vs host galaxy absolute magnitude relationship, which can be used for establishing the mass once the
host galaxy luminosity is measured. 

\begin{figure}[h]
\centering
\includegraphics[width=1.0\columnwidth]{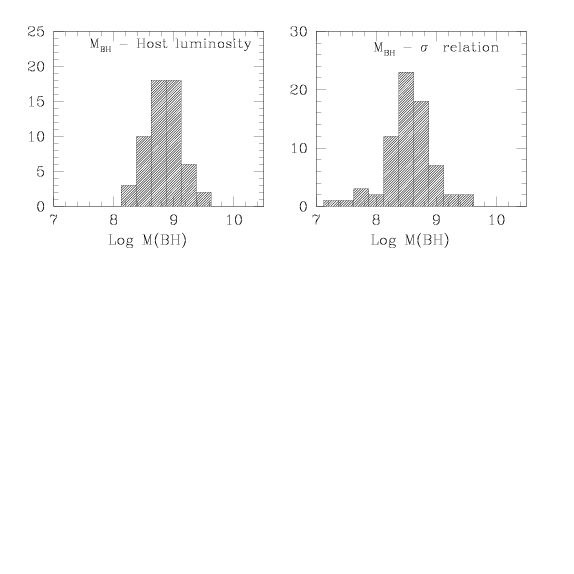}
\vspace{-6cm}
\caption{Left panel: distribution of black hole masses  of BLLs  derived from the luminosity of their  host galaxies  \citep{falomo2003b}.
Right panel: distribution of black hole masses from  the $\sigma - M_{BH}$ relation 
\citep{plotkin2011}.}
\label{mbh_bll}
\end{figure}

In the case of BLLs the main difficulty is to measure the luminosity of the host
galaxy (see Section \ref{sec:hostgals}). From systematic measurements of the host galaxies luminosity of 
BLL up to $z \sim 0.5$ the black hole  masses have been
calculated by  \citet{woo2002} and \citet[][see Fig. \ref{mbh_bll}]{falomo2003b}. Beyond that
redshift limit the host galaxies become very difficult to detect, and evolutionary effects  may become
important, requiring corrections to the relations obtained locally. 

Another parameter  of the host galaxy, which is related to the mass of the black hole is the
velocity dispersion $\sigma$  of the central part of the galaxy. It is measured through the width
of the absorption lines (typically $H_\beta$)  of the stellar component. In terms of dispersion
the $\sigma-M_{BH}$ relation is somewhat preferable  than that discussed in the previous paragraph.
The technique  for  mass determination of BLLs  was introduced by  \cite{falomo2002}, in this way 
it was determined the mass of a dozen objects \cite{falomo2003a}  and \cite{barth2003}. 
It was extended to $\sim$60 objects by  \cite{plotkin2011}, who considered the sample  of optically selected BLL
candidates from the SDSS DR7  (see Section \ref{sec:demograph} and  Fig. \ref{mbh_bll}). Also in this case
there is a maximum value of $z \sim0.5$, for the applicability of the method, related to the
possibility of having reliable measurements of $\sigma$. 
The mass distributions reported in Fig. \ref{mbh_bll} are very similar to those of  quasars, as deduced 
from the width of broad lines and the continuum intensity  \citep[see e.g.][]{shen2011}.

Finally another technique, mutuated from X-ray binaries, is based on Shakura-Sunyaev disk model
\citep{shakura1973}.  The idea is to 
consider the spectral energy distribution of the source in a broad frequency range, exclude the
contribution of the jet, and isolate the one from the accretion disk. In practice one should be
able to observe the region where $\nu F_{\nu}$ has a maximum. The fit with a  Shakura-Sunyaev disk
allows one to constrain the mass and the accretion rate. If the redshift is known the latter quantity can be fixed
and one obtains the mass. Up to now, results have been obtained for high redshift blazars (z $> 2$), rather
than BLLs \citep{sbarrato2013}.


\section{Concluding Remarks}
\label{sec:conclrem}

About half a century after the discovery of the first BLLs the basic picture to interpret this enigmatic class 
of sources appears robust: BLLs are active nuclei of massive elliptical galaxies  the emission of which is  dominated by  relativistic jets. The bright compact radio cores, high luminosities and rapid, large amplitude flux variability at all frequencies and the strong  polarization that characterise these objects are well explained in this scenario.
Also the observed quasi featureless optical spectra of BLL, that is one of the distinctive properties of the class, find a natural 
explanation in the above beaming model. At variance with other classes of AGN, like quasars and Seyfert galaxies, 
the lack of prominent spectral features was, and still in part is, a significant limiting factor for the determination of their redshift and  ensuing evaluation of their physical properties.
The  implication of this model is the existence of a large number of misaligned objects with the same intrinsic properties  as BLLs. The most  obvious candidates are low-power radiogalaxies.

BLLs were mainly discovered as counterparts of radio and/or X-ray sources.
The surveys in these bands led to the compilation of the first, albeit relatively small, complete samples of objects from which 
it was possible to  start the exploration of  
the cosmic evolution of the class as compared with other types of AGN. Because of the limited number of known 
BLL and the paucity of known objects at high  redshift  ($z \gs 1$) the evolution with cosmic time remains vague.  Negative or positive evolution were suggested for different samples, although almost consistent with no evolution, yielding a picture substantially different from that of the majority of AGNs, where significant positive evolution is seen.

The investigation of the cosmic evolution of BLLs is an important issue that could offer relevant clues on the relationship between the formation of supermassive black holes and the development of relativistic jets.  In order to achieve a significant progress  in the understanding of the cosmic evolution of BLLs  the extension of available samples to higher redshift is mandatory.

The discovery and follow-up study of high-redshift BLLs will come within reach thanks to the new generation of observing facilities at all frequencies foreseen for the next decades (e.g., Square Kilometer Array, Extremely Large Telescopes, James Webb Space Telescope, Large Synoptic Survey Telescope, Advanced Telescope for High Energy Astrophysics, Cherenkov Telescope Array).  In particular,
high signal-to-noise ratio spectroscopy of BLL candidates obtained with adaptive optics instrumentation at  extremely large telescopes  in the near-infrared   will allow one to 
observe the rest-frame optical spectrum of high-redshift sources and determine the redshift from absorption features of the host galaxies or faint emission lines. This will produce the first sizable high-redshift sample of BLLs, that, compared with the lower redshift samples,  will yield a firm conclusion on cosmic evolution of the class.   This could be related to the  role of the  angular momentum of the central black hole  in forming a relativistic jet.

Although the observational knowledge of the BLL phenomenon is relatively well established,  on the theoretical front some fundamental aspects remain unsolved. Specifically, a  central problem of BLL physics is the mechanism for the extraction and transfer of energy from the inner engine to the jets.   
We trust that the observations with the above-mentioned future facilities will provide huge gain in terms of signal-to-noise ratio, spatial and time resolution for spectroscopy, photometry and polarimetry.  
Some examples of the expected capabilities for the determination of the redshift with future optical and near infrared facilities are given in \cite{landoni2013} and 
\cite{landoni2014}.

These crucial measurements will  also contribute to the comprehension of the mechanisms  of conversion of rotational energy into kinetic energy of the jet and into luminosity in various channels (electromagnetic, neutrinos, gravitational radiation). 

\begin{acknowledgements}
We are grateful to  Catherine Boisson, Alessandro Caccianiga, Stefano Covino, Luigi Foschini,  Gabriele Ghisellini, Oliver King, Alan Marscher, Claudia Raiteri, Kenji Toma, Massimo Villata, Andreas Zech for inputs in the preparation of this work.   In particular, we thank Gabriele Giovannini and Massimo Persic for useful suggestions, comments and discussions on this review.  We are also very grateful to Angela Sandrinelli and Marco Landoni for a  careful reading of the manuscript and for fruitful comments.  We acknowledge support from ASI/INAF grant  I/088/06/0 and PRIN MIUR 2010/2011.
\end{acknowledgements}

\end{document}